\documentclass[onecolumn,prd,amssymb,amsmath,preprintnumbers,aps,nofootinbib,superscriptaddress,11pt]{revtex4-1}
\usepackage[a4paper, portrait, margin=0.7in]{geometry}
\usepackage{bm,color,xcolor}
\usepackage{slashed} 
\usepackage{graphicx}
\usepackage{soul} 
\usepackage{multirow}
\usepackage{makecell}
\usepackage{comment}
\usepackage{mathtools}
\usepackage{draftfigure}
\usepackage{cancel}
\usepackage{appendix}
\usepackage[colorlinks=true
,urlcolor=blue
,anchorcolor=blue
,citecolor=blue 
,filecolor=blue
,linkcolor=blue
,menucolor=blue
,linktocpage=true
,pdfproducer=medialab
,pdfa=true
]{hyperref}

\newcommand{\MeV}{\ensuremath{\mathrm{MeV}}}
\newcommand{\GeV}{\ensuremath{\mathrm{GeV}}}
\newcommand{\TeV}{\ensuremath{\mathrm{TeV}}}

\newcommand{\beq}{\begin{equation}}
\newcommand{\eeq}{\end{equation}}
\newcommand{\bea}{\begin{eqnarray}}
\newcommand{\eea}{\end{eqnarray}}

\begin{document}
\title{\boldmath \bf \Large 
Enhancing Searches for Heavy QCD Axions via Dimuon Final States
}
\preprint{UMN-TH-4201/22, FTPI-MINN-22/25}

\author{Raymond T. Co}
\email{rco@umn.edu}
\thanks{\scriptsize \!\! \href{https://orcid.org/0000-0002-8395-7056}{0000-0002-8395-7056}}
\affiliation{\it School of Physics and Astronomy, University of Minnesota, Minneapolis, MN 55455, USA}
\affiliation{\it William I. Fine Theoretical Physics Institute, University of Minnesota, Minneapolis, MN 55455, USA}
\author{Soubhik Kumar}
\email{soubhik@berkeley.edu}
\thanks{\scriptsize \!\! \href{https://orcid.org/0000-0001-6924-3375}{0000-0001-6924-3375}}
\affiliation{Berkeley Center for Theoretical Physics, Department of Physics, University of California, Berkeley, CA 94720, USA}
\affiliation{Theoretical Physics Group, Lawrence Berkeley National Laboratory, Berkeley, CA 94720, USA}

\author{Zhen Liu}
\email{zliuphys@umn.edu}
\thanks{\scriptsize \!\! \href{https://orcid.org/0000-0002-3143-1976}{0000-0002-3143-1976}}
\affiliation{\it School of Physics and Astronomy, University of Minnesota, Minneapolis, MN 55455, USA}

\begin{abstract}
Heavy QCD axions are well-motivated extensions of the QCD axion that address the quality problem while still solving the strong {\it CP} problem. Owing to the gluon coupling, critical for solving the strong {\it CP} problem, these axions can be produced in significant numbers in beam dump and collider environments for axion decay constants as large as PeV, relevant for addressing the axion quality problem. In addition, if these axions have leptonic couplings, they can give rise to long-lived decay into lepton pairs, in particular, dominantly into muons above the dimuon threshold and below the GeV scale in a broad class of axion models. Considering existing constraints, primarily from rare meson decays, we demonstrate that current and future neutrino facilities and long-lived particle searches have the potential to probe significant parts of the heavy QCD axion parameter space via dimuon final states.
\end{abstract}
{
\let\clearpage\relax
\maketitle
}

\tableofcontents
\section{Introduction}

The quantum chromodynamics (QCD) axion was proposed to address the strong {\it CP} problem~\cite{tHooft:1976rip}, which concerns the fact that the {\it CP} violation in the strong interactions is experimentally constrained to be less than $\mathcal{O}(10^{-10})$ via the non-observation of neutron electric dipole moment~\cite{Baker:2006ts, Afach:2015sja, Graner:2016ses, Abel:2020gbr} as opposed to the theoretical expectation of $\mathcal{O}(1)$~\cite{tHooft:1976rip}. 
In the Peccei-Quinn (PQ) mechanism~\cite{Peccei:1977hh,Peccei:1977ur}, the QCD axion~\cite{Weinberg:1977ma,Wilczek:1977pj} is coupled to the gluons, and upon confinement the QCD dynamics generates a QCD axion potential with a {\it CP}-conserving minimum and a mass $m_a \simeq 5.7~{\rm meV}\times(10^9~\GeV / f_a)$~\cite{Borsanyi:2016ksw,Gorghetto:2018ocs} with $f_a$ being the axion decay constant. 
As the QCD axion relaxes to this minimum, the strong {\it CP} problem is solved dynamically.

While this is an elegant mechanism to address the strong {\it CP} problem, in the minimal realization it suffers from the axion quality problem~\cite{Holman:1992us,Barr:1992qq,Kamionkowski:1992mf,Dine:1992vx, Ghigna:1992iv}.
To illustrate this, we can model the axion $a$\footnote{From now on, we will use the phrase `axion' to denote both the standard QCD axion and its heavier variants. The phrase `axion-like particles'~(ALPs), on the other hand, will be reserved for pseudoscalars not addressing the strong {\it CP} problem, as often done in the current literature.} as a (pseudo) Nambu-Goldstone boson residing in the PQ field $\Phi  \sim f_a e^{i a/f_a}$, which is charged under a global and anomalous $U(1)_{\rm PQ}$ symmetry.
Since gravitational effects are expected to break global symmetries~\cite{Perry:1978fd,Hawking:1979hw,Giddings:1988cx,Gilbert:1989nq,Kallosh:1995hi,Harlow:2018jwu,Harlow:2018tng}, including $U(1)_{\rm PQ}$, we expect Planck scale suppressed terms such as ${\cal L}\supset \Phi^n/M_{\rm Pl}^{n-4}$ to arise.
Written in terms of the axion, we then have ${\cal L}\supset f_a^n \cos(na/f_a + \varphi_n) / M_{\rm Pl}^{n-4}$ where $\varphi_n$ is the complex phase of the coefficient of this term.
For $n > 4$, these contributions can potentially drive the axion away from the {\it CP}-conserving minimum due to the random nature of the phases $\varphi_n$, and consequently spoil the solution to the strong {\it CP} problem.
The situation is exacerbated in minimal scenarios with large values of $f_a$, as is the case with the conventional QCD axion. 
This scenario concerns $m_a \lesssim \mathcal{O}(10)$ meV, which is very sensitive to the above corrections.
Even with the lowest decay constant $f_a$ allowed by astrophysical bounds~\cite{{Ellis:1987pk,Raffelt:1987yt,Turner:1987by,Mayle:1987as,Raffelt:2006cw,Payez:2014xsa,Chang:2018rso,Carenza:2019pxu,Bar:2019ifz,Buschmann:2021juv}}, $f_a \simeq 10^8~\GeV$, all Planck-suppressed operators up to dimension-8 have to be severely constrained to avoid shifting the potential minimum by $\mathcal{O}(10^{-10})$.
It is then vital to understand why the PQ symmetry is of such a high {\it quality}.

This problem is significantly relaxed if the gravitational corrections are suppressed and/or if the {\it CP}-conserving potential is strengthened so that the potential is more stable against {\it CP}-violating corrections. 
The former is achieved by $f_a \ll 10^8~\GeV$, which then requires $m_a \gtrsim 100$~MeV to avoid various astrophysical bounds. 
The latter is the case if the axion mass is much larger than that dictated by the strong dynamics---as in the so-called {\it heavy} QCD axion models. 
The axion mass can be enhanced in ways that still preserve the {\it CP} symmetry~\cite{Dimopoulos:1979pp, Rubakov:1997vp,Berezhiani:2000gh,Hook:2014cda,Fukuda:2015ana,Gherghetta:2016fhp, Dimopoulos:2016lvn, Agrawal:2017ksf, Agrawal:2017evu, Gaillard:2018xgk, Lillard:2018fdt, Csaki:2019vte, Hook:2019qoh, Gherghetta:2020keg, Gherghetta:2020ofz, Valenti:2022tsc,Kivel:2022emq}. (See also~\cite{Tye:1981zy, Holdom:1982ex, Flynn:1987rs} for early work on raising the axion mass.) 
In this regime inspired by the axion quality problem where $m_a \gtrsim 100~\MeV$ and $f_a \ll 10^8~\GeV$, the heavy QCD axions are more strongly interacting with the Standard Model and can be produced and searched for in the collider and beam dump experiments. In this regard, various approaches have been pursued at beam dump, flavor, and collider experiments~\cite{Essig:2010gu, Dobrich:2015jyk, Dolan:2017osp, Dobrich:2019dxc, Harland-Lang:2019zur, Dent:2019ueq, Brdar:2020dpr, Jaeckel:2012yz, Mimasu:2014nea, Jaeckel:2015jla, Izaguirre:2016dfi, Knapen:2016moh, Bauer:2017ris, Brivio:2017ije, Mariotti:2017vtv,  CidVidal:2018blh, Beacham:2019nyx, Alonso-Alvarez:2018irt, Ebadi:2019gij, Gavela:2019cmq, Gavela:2019cmq, Altmannshofer:2019yji, Gershtein:2020mwi, Knapen:2021elo}, both for heavy QCD axions and more generally ALPs.
Furthermore, heavy QCD axions or ALPs can also play important roles in astrophysics and cosmology, such as explaining the dark matter and baryon abundance~\cite{Cohen:1987vi, Nomura:2008ru, Co:2019wyp, Domcke:2020kcp, Co:2020jtv, Co:2022aav, Foster:2022ajl, Panci:2022wlc}.

In addition to the defining gluon coupling, the axion may couple universally to all the other Standard Model (SM) gauge bosons as predicted by grand unification and also to the SM fermions in a broad class of theories, including the DFSZ models~\cite{Dine:1981rt,Zhitnitsky:1980tq}, or to new heavy quarks as in the KSVZ models~\cite{Kim:1979if,Shifman:1979if}.
The coupling to fermions implies that the axion may dominantly decay into a pair of fermions when kinematically allowed, opening the possibility of unique experimental signatures. 
Specifically, in this work, we propose a search for heavy QCD axions, with masses above the dimuon threshold and below the GeV scale, at various neutrino and beam dump experiments.
For these masses, axions may dominantly decay into a pair of muons, as we will describe in detail below.
As examples, we focus on neutrino experiments utilizing
the liquid argon time projection chamber (LArTPC)~\cite{Rubbia:1977zz} technology, such as the Short-Baseline Near Detector (SBND)~\cite{MicroBooNE:2015bmn}, ICARUS~\cite{MicroBooNE:2015bmn} and Deep Underground Neutrino Experiment (DUNE)~\cite{DUNE:2018hrq}. 
The dimuon final state can be particularly useful from the background mitigation perspective, especially after applying an invariant mass cut. 
While the dielectron final state can also be interesting, for the benchmark models that we consider, the branching ratio to dielectrons is subdominant compared to that into diphotons.
Above the GeV range, axions would predominantly decay hadronically and constitute a different class of signatures explored in~\cite{Kelly:2020dda} in the context of DUNE. 
Therefore we focus on the mass range between the dimuon threshold and $\mathcal{O}$(GeV). 
Through the gluon coupling, the axions are produced via its mixing with SM mesons produced when the beams hit the target or the absorber. The axions can then propagate to and decay within the LArTPC of the experiments, where the muons will leave two distinct minimally ionizing tracks. 
We also perform similar analyses with long-lived particle searches in the context of SHiP~\cite{SHiP:2015vad} and FASER~2~\cite{Feng:2017uoz,FASER:2019aik,Feng:2022inv}.

Recently, such a search has been performed for the ArgoNeuT detector~\cite{Anderson:2012vc} using data collected in 2009-2010 in the Neutrinos at the Main Injector (NuMI) beamline~\cite{Adamson:2015dkw} at Fermilab, and an important constraint in the axion parameter space is obtained in the mass range $m_a$ between 0.2-0.9~GeV for an axion decay constant $f_a$ around 10-100 TeV~\cite{ArgoNeuT:2022mrm}. 
The dimuon signatures have also been exploited in Refs.~\cite{LHCb:2015nkv, LHCb:2016awg} for LHCb and Ref.~\cite{Dobrich:2018jyi} for CHARM, respectively, where axions are assumed to be produced from the coupling with the top quark instead. 
Similarly, utilizing LArTPC but assuming the absence of the axion-fermion coupling, Ref.~\cite{Kelly:2020dda} analyzes the sensitivity of the DUNE detector with the gluon coupling, and Ref.~\cite{Brdar:2020dpr} shows prospects for a DUNE-like detector without the gluon coupling.

\begin{figure}
    \centering
    \includegraphics[width = 0.6\textwidth]{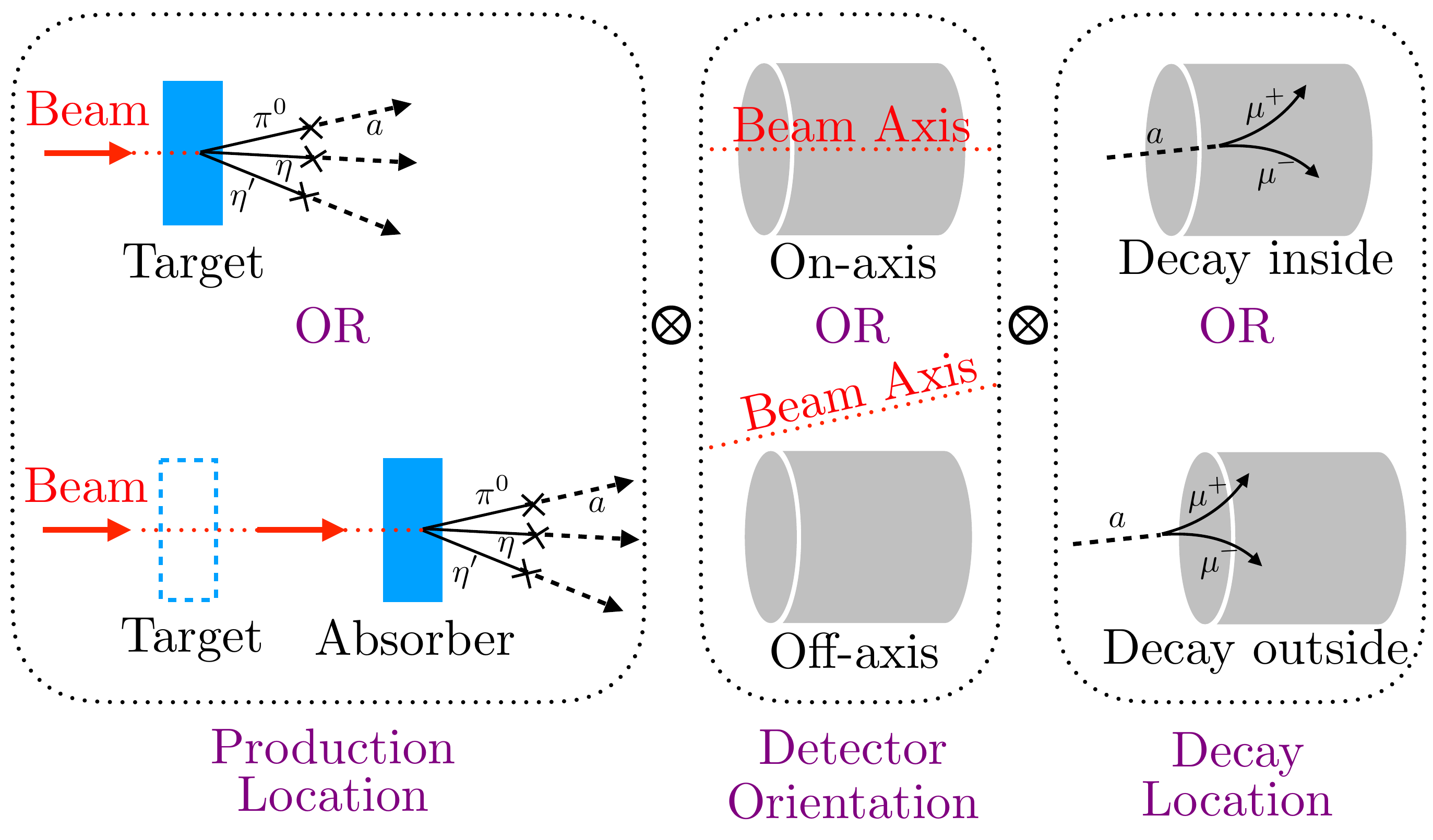}
    \caption{In this schematic diagram, the three columns show possible production locations, beam directions with respect to the detector, and decay locations. The axions can be produced from the beam at the target or at the absorber. The detector, shown by the gray cylinder, may be on- or off-axis from the beam. Lastly, the axion may decay inside or before entering the detector.}
    \label{fig:schematic}
\end{figure}

We illustrate the various experimental setups in Fig.~\ref{fig:schematic}. Three aspects characterize this.
Firstly, the axion production can be dominated by either the target or the absorber located further downstream.
While the absorber would receive less flux compared to the target, due to the proximity of the absorber to the detector, it can dominate the experimental sensitivity. 
This was found to be the case for the ArgoNeuT search in Ref.~\cite{Anderson:2012vc}. 

Secondly, while each detector is typically on the same axis as its associated beam line, sometimes a detector can be more sensitive if axions produced from a separate, simultaneously operating off-axis beam reach it~\cite{deGouvea:2018cfv, Batell:2019nwo}.
As an example, while the ICARUS detector is nominally associated and on the same axis with the 8~GeV Booster Neutrino Beam (BNB), due to its large volume it can receive a large flux of axions produced as the 120~GeV protons at the NuMI beam hit the NuMI target, even if the NuMI beam axis does not pass through ICARUS directly.
This increased sensitivity to NuMI beam compared to BNB has to do with the fact that the 120~GeV NuMI beam produces a larger flux of axions and also the fact that ICARUS is not too off-axis to lose that flux.
We will make a quantitative comparison between the results with the two beams in Sec.~\ref{sec:bd}.

Thirdly, since our search strategy involves a dimuon final state and the produced muons from axion decay are often very energetic, they can penetrate the earth/material before the detector. 
Thus to consider such events, we include axion decays both inside and outside of the detector.
This increase in effective decay volume can have an important effect on the experimental sensitivity, as we will illustrate in the context of DUNE near detector in Sec.~\ref{sec:bd}.
To give another example in this context, in the ArgoNeuT search~\cite{Anderson:2012vc}, we considered an extra decay length of 63~cm before the detector front panel.

This work is organized as follows. 
In Sec.~\ref{sec:axion_EFT} we review the axion quality problem and motivate how heavy QCD axions can improve it.
Then we describe the couplings of the heavy QCD axion to the SM using an effective field theory (EFT) framework.
After describing the various decay modes of the axion, in Sec.~\ref{sec:bd} we study the details of axion production in various neutrino and beam dump experiments. 
With these results at hand, we derive the projected reach that SBND, ICARUS, DUNE, FASER~2, and SHiP may be able to achieve, and then summarize the existing constraints on the axion parameter space coming mostly from rare meson decays.
In Sec.~\ref{sec:UV} we describe examples of UV completions that can give rise to the axion EFT under consideration.
We conclude in Sec.~\ref{sec:concl}.

\section{Heavy QCD Axion EFT}\label{sec:axion_EFT}
We start our general analysis with an EFT approach, and we will present examples of UV realization in Sec.~\ref{sec:UV}.
After reviewing the role the heavy axions play in the context of the quality problem, we describe the EFT and summarize the decay modes of the axion.

\subsection{Axion Quality Problem and Heavy QCD Axions}
As alluded to in the introduction, the axion quality problem is that $U(1)_{\rm PQ}$ breaking contributions could generally arise from gravitational corrections.
Such corrections can give rise to new {\it CP} non-conserving minima, and as the axion dynamically relaxes to such a minimum, the strong {\it CP} problem reappears.
To be quantitative, we first consider the case of the standard QCD axion which couples to QCD as
\begin{align}
    \mathcal{L} \supset \frac{\alpha_s}{8\pi} \left( \frac{a}{f_a} + \bar{\theta}\right) G_{\mu\nu}^a \tilde{G}^{a,\mu\nu}.
\end{align}
Upon QCD confinement this gives rise to an axion potential~\cite{DiVecchia:1980yfw,GrillidiCortona:2015jxo}
\begin{align}\label{eq:QCDpot}
    V_{\rm QCD}(a) = -m_\pi^2 f_\pi^2 \sqrt{1 - \frac{4m_u m_d}{(m_u+m_d)^2}\sin^2\left(\frac{a}{2 f_a}+\frac{\bar{\theta}}{2}\right)}.
\end{align}
Here $m_\pi, f_\pi$ are the pion mass and decay constant, and $m_u, m_d$ are up and down quark masses, respectively.
This determines the QCD axion mass to be,
\begin{align}\label{eq:qcd_axion_mass}
    m^{\rm QCD}_a \approx \frac{\sqrt{m_u m_d}}{m_u+m_d}\frac{m_\pi f_\pi}{f_a} \approx 5.7 \left(\frac{10^9~{\rm GeV}}{f_a}\right)~{\rm meV}.
\end{align}

To see how the axion quality problem arises in this context, consider the case where the axion is a pseudo-Nambu-Goldstone boson of a spontaneously broken $U(1)_{\rm PQ}$ symmetry.
The axion can then be written as the phase residing in the PQ scalar, $\Phi  \sim f_a e^{ia / f_a}$.
Gravitational violation of this global PQ symmetry effects would then imply Lagrangian terms like 
\begin{align}\label{eq:gravity}
    \frac{\Phi^n}{M_{\rm Pl}^{n-4}} \supset \frac{f_a^n}{M_{\rm Pl}^{n-4}} \cos\left(\frac{n a}{f_a} + \varphi_n\right)
\end{align}
would be present, and they would give rise to extra contributions to the axion potential.
Now a variety of astrophysical constraints on the QCD axion imply $f_a \gtrsim 10^8$~GeV~\cite{Ellis:1987pk,Raffelt:1987yt,Turner:1987by,Mayle:1987as,Raffelt:2006cw,Payez:2014xsa,Chang:2018rso,Carenza:2019pxu,Bar:2019ifz,Buschmann:2021juv}.
Given the light axion mass dictated by Eq.~\eqref{eq:qcd_axion_mass}, we then see that unless we forbid all higher dimensional operators up to $n=8$, the gravitational contributions would move the axion away from the {\it CP} conserving minima, spoiling the solution to the strong {\it CP} problem.

To contrast this scenario, we now focus on heavy QCD axions.
As an example for this class of models, consider the scenario where the axion couples to a dark gauge group that confines at a scale $\Lambda_D \gg \Lambda_{\rm QCD}$.
As a result, the axion obtains a potential of the type,
\begin{align}\label{eq:heavy_ax_pot}
    V_D(a) \simeq \Lambda_D^4 \cos\left(\frac{a}{f_a}+\delta\right).
\end{align}
Due to the fact that $\Lambda_D \gg \Lambda_{\rm QCD}$, the dynamical relaxation of the axion is mostly dictated by this potential, rather than Eq.~\eqref{eq:QCDpot}.
Because of the structure of the theory, such as the presence of an additional $\mathbb{Z}_2$ mirror symmetry~\cite{Rubakov:1997vp, Berezhiani:2000gh, Hook:2014cda, Fukuda:2015ana, Dimopoulos:2016lvn, Hook:2019qoh} or embedding of $SU(3)_c$ into larger gauge groups~\cite{Dimopoulos:1979pp, Gherghetta:2016fhp, Gaillard:2018xgk, Agrawal:2017evu, Gherghetta:2020ofz, Valenti:2022tsc}, the axion can still solve the strong {\it CP} problem while being heavier than the QCD axion.

Now the situation is quite improved from the perspective of the quality problem.
Through Eq.~\eqref{eq:heavy_ax_pot}, the axion mass is given by $m_a \simeq \Lambda_D^2/f_a$ and is much larger than that from Eq.~(\ref{eq:QCDpot}) for identical values of $f_a$.
Existing constraints on such heavier axions can be much weaker compared to the QCD axion.
In particular, as the axion mass becomes larger than ${\cal O}(100)$~MeV, the astrophysical and cosmological constraints can weaken and parameter space with $f_a\ll 10^8$~GeV is then allowed by the current set of constraints. 
Consequently, these smaller values of $f_a$ reduce the strength of the gravitational contributions seen in Eq.~\eqref{eq:gravity} and correspondingly, the axion potential becomes robust against such corrections.
In particular, for some of the $m_a$ and $f_a$ values we consider in the rest of this work, one needs to forbid only the dimension-5 term in Eq.~\eqref{eq:gravity}.
This, therefore, reduces the severity of the quality problem.

\subsection{Lagrangian and Low Energy Effective Theory}
Generic ALPs can couple to the SM in a variety of ways. However, the heavy QCD axions that we focus on in this work have a defining coupling to QCD since that is essential for solving the strong {\it CP} problem. At the same time, motivated by gauge coupling unification, we focus on a class of EFTs where axions also couple to $SU(2)_L$ and $U(1)_Y$ gauge bosons. Therefore, we consider the following gauge sector coupling,
\begin{align}
\label{eq:lag_gauge}
\mathcal{L}_{\rm gauge} = c_3\frac{\alpha_s}{8\pi f_a}a G\tilde{G} + c_2\frac{\alpha_2}{8\pi f_a}a W\tilde{W} + c_1\frac{\alpha_1}{8\pi f_a}a B\tilde{B}.
\end{align}
This particular set of couplings also appears in the context of KSVZ axions. In the following, we will set $c_3=c_2=c_1=1$. Along with the gauge coupling, axions can also couple to fermions. In the present work, we consider only flavor-diagonal lepton couplings of the axion. The motivation is two-fold. First, since the axion does not couple to quarks and also does not give rise to flavor-changing processes at the tree level, we can focus on a parameter space that is not strongly constrained by existing searches, such as those of rare meson decays (see, {\it e.g.},~\cite{Goudzovski:2022vbt} for a recent summary). This then lets us focus on a complementary part of the parameter space. Second, given the leptonic coupling of the axion, it can dominantly decay into a pair of muons. As we will discuss later, the resulting dimuon signature can be a powerful discovery channel for GeV-scale axions. 
Therefore, we consider a lepton sector coupling,
\begin{align}\label{eq:lag_lepton}
\mathcal{L}_{\rm lepton} = \sum_{\ell=e,\mu,\tau} c_{\ell} \frac{\partial_\mu a}{2f_a}\bar{\ell}\gamma^{\mu}\gamma_5\ell.
\end{align}
As we will show in Sec.~\ref{sec:UV}, such a scenario can naturally arise in UV complete models. To summarize, the axion EFT in our scenario is given by,
\begin{align}
	\mathcal{L}_{\rm axion} =
	\frac{1}{2}(\partial_\mu a)^2 - \frac{1}{2}m_a^2 a^2 + 
	\mathcal{L}_{\rm gauge} + \mathcal{L}_{\rm lepton},
\end{align}
at some UV scale $\Lambda \sim 4\pi f_a$. 
Below $\Lambda$, various other operators are introduced via renormalization group (RG) evolution. We follow Refs.~\cite{Bauer:2021wjo, Bauer:2021mvw} to account for the RG evolution and compute the axion couplings below the GeV scale.

\subsection{Axion Decay}
Given the EFT in Eq.~\eqref{eq:lag_gauge} and Eq.~\eqref{eq:lag_lepton}, the axion dominantly decays into three classes of final states: ($a$) diphotons, ($b$) dimuons, and ($c$) hadrons. We discuss each of them briefly now.
\paragraph{Photons:} The width for an axion decaying into two photons is given by,
\begin{align}
\Gamma_{a\rightarrow \gamma\gamma} = \frac{\alpha^2 |c_\gamma|^2 m_a^3}{256\pi^3 f_a^2}.
\end{align}
Here $c_\gamma$ is determined in terms of the EFT coefficients appearing in Eqs.~\eqref{eq:lag_gauge} and \eqref{eq:lag_lepton} as~\cite{Bauer:2017ris, Aloni:2018vki, Georgi:1986df},
\begin{equation}
\begin{aligned}
c_\gamma & = c_3\left(-1.92+\frac{1}{3}\frac{m_a^2}{m_a^2-m_\pi^2}+\frac{8}{9}\frac{m_a^2-\frac{4}{9}m_\pi^2}{m_a^2-m_\eta^2}+\frac{7}{9}\frac{m_a^2-\frac{16}{9}m_\pi^2}{m_a^2-m_{\eta'}^2}\right) + \frac{5}{3} c_1+ c_2 + 2 \sum_{\ell=e,\mu,\tau} c_{\ell} B(4m_\ell^2/m_a^2),
\label{eq.cgalow}
\end{aligned}
\end{equation}
and $\alpha$ the fine-structure constant.
Here $B(x) = 1- x  f(x)^2$ and 
\begin{align}
f(x)=
\begin{cases}
\sin^{-1}\left(\frac{1}{\sqrt{x}}\right)    & \quad \text{if }  x\geq 1\\
    \frac{\pi}{2} + \frac{i}{2} \log\left(\frac{1+\sqrt{1-x}}{1-\sqrt{1-x}}\right) & \quad \text{if } x<1
\end{cases} \, .
\end{align}
For our choices of $c_i$, the diphoton mode dominates below the dimuon threshold.

\paragraph{Muons:}
The axion width into muons is given by,
\begin{align}
\Gamma_{a\rightarrow \mu\mu} = \frac{c^2_{\ell} m_a m_\mu^2}{8\pi f_a^2} \sqrt{1-\frac{4m_\mu^2}{m_a^2}}.    
\end{align}
Here and in the following, we will use $c_\ell$ to denote all the lepton couplings, {\it i.e.}, assuming them to be flavor-diagonal. Soon above the kinematic threshold, the muon decay modes dominate over the diphoton channel, except in regions where the axion resonantly mixes with SM pseudoscalar mesons.

\paragraph{Hadrons:}
Below the QCD scale, the axion also couples to various hadrons, thanks to the axion-gluon coupling. For $m_a\lesssim$~GeV, the dominant decay channels are into $\pi\pi\pi, \eta\pi \pi, \pi\pi\gamma$. We use the results of Ref.~\cite{Aloni:2018vki} to take these decays into account.
The relative importance of these modes compared to the dimuon mode depends on the values of $c_\ell$, and we consider two benchmark scenarios $c_\ell = 1/36$ and $c_\ell = 1/100$ to illustrate the differences. 
As will be discussed in more detail in Sec.~\ref{sec:UV}, the coefficient $c_\ell$ roughly corresponds to a mixing angle squared $\theta_{\rm mix}^2$ in one UV completion that we will focus on.
Thus the choices $c_\ell = 1/36, 1/100$ originate from $\theta_{\rm mix}\simeq 1/6, 1/10$, respectively.

\begin{figure}
	\includegraphics[width=0.47\textwidth]{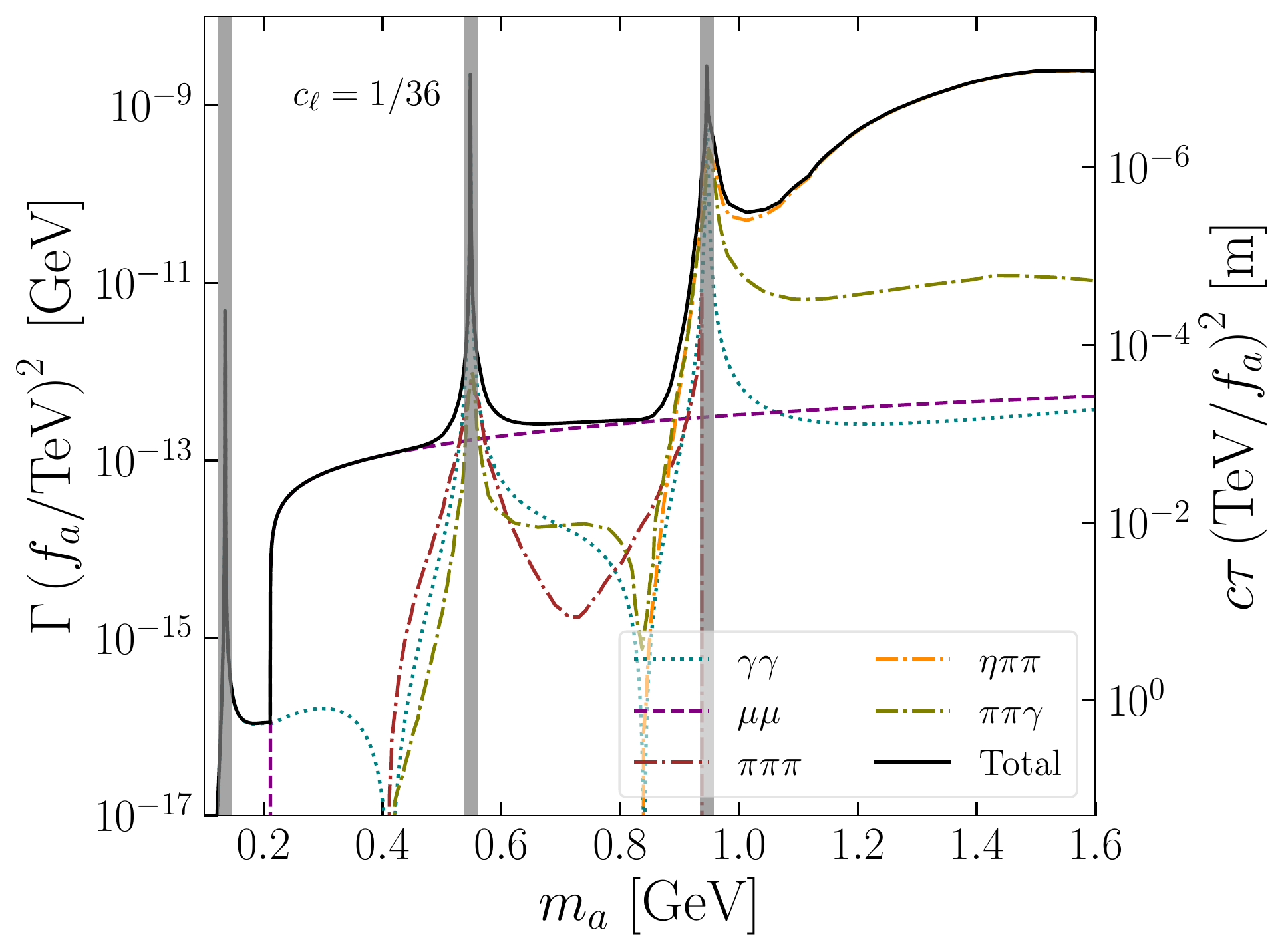}
	\includegraphics[width=0.47\textwidth]{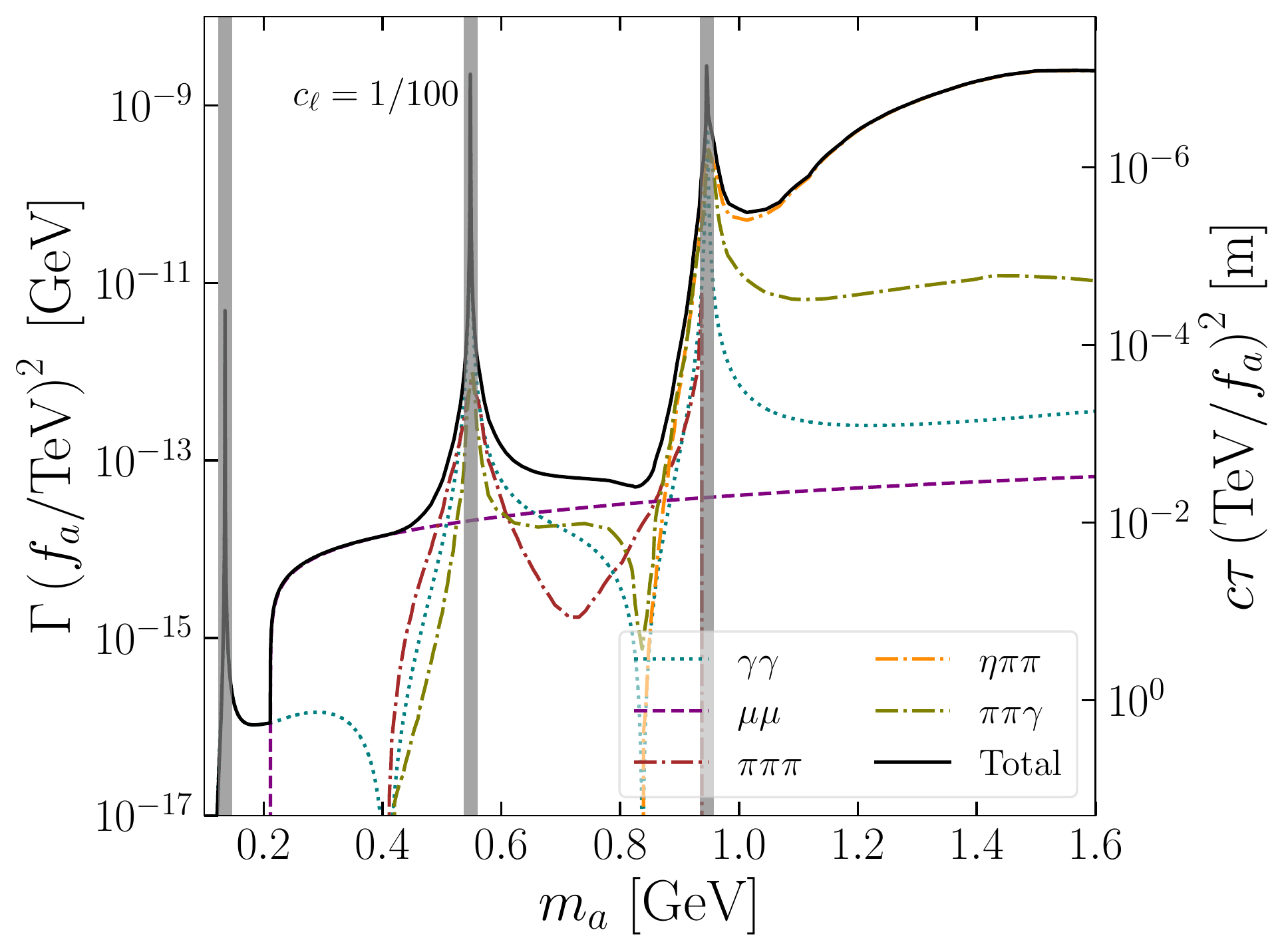}
	\caption{The relative contributions of various decay channels to the total width of the axion for $c_\ell = 1/36$ (left) and $c_\ell = 1/100$ (right). For $2m_\mu < m_a \lesssim 0.8$~GeV, the axion dominantly decays into a pair of muons.}
	\label{fig:partial_width}
\end{figure}

The relative contributions to the decay width of the photon mode (dotted), muon mode (dashed), and hadronic modes (dot-dashed) for the two benchmark choices are shown in Fig.~\ref{fig:partial_width}.

\section{Beam Dump Considerations}\label{sec:bd}
In this section, we discuss some of the details of axion production in a beam dump environment in the presence of a gluon coupling. We consider several neutrino experiments: ArgoNeuT~\cite{Anderson:2012vc}, SBND~\cite{MicroBooNE:2015bmn}, ICARUS~\cite{MicroBooNE:2015bmn} and DUNE~\cite{DUNE:2018hrq} for which axions are produced when a proton beam hits a stationary target (or an absorber placed downstream). While a similar setup is also true for SHiP~\cite{SHiP:2015vad}, we also study the discovery capabilities of FASER~2 for which axions would get produced in 14~TeV $pp$ collisions at the LHC. The axions produced in the forward direction can then travel to the FASER~2 detector.

\subsection{Production}\label{sec:production}
The theory of axion production in the presence of gluon couplings is an active area of research. Along with axion production in meson mixing or gluon fusion for higher masses, axion production in proton bremsstrahlung can also be an important contribution. For scalars, this was computed in Refs.~\cite{Boiarska:2019jym, Foroughi-Abari:2021zbm,SplittingAxions}, while for the case of pseudoscalars, there are subtleties. These subtleties~\cite{SplittingAxions} include the axion flux dependence 
on the exact form of quasi-real approximation, the interference between initial state radiation and final state radiation, the comparison of the proton bremsstrahlung calculation and the hadronization contribution in the regime of concern, as well as the exact order of mass expansion and momentum expansion. Some of these subtitles arise because the phase space of interest involves scales comparable to the axion mass and the {\it CP} nature of the axion-fermion coupling, and because the formalism itself is only an approximation heavily dependent on the matching procedure and scheme at a fixed order. Given all these subtleties, we still approximate the axion production rate by the SM pseudoscalar production rate multiplied by the square of the axion-pseudoscalar mixing angle, as is widely done in the current literature.
Since these mixing angles depend on the axion-quark coupling, a field rotation-dependent quantity~\cite{Blinov:2021say,SplittingAxions}, such an approximation needs to be improved.
Furthermore, this approximation also heavily relies on the SM QCD production of mesons, and involves an unphysical, ad hoc assignment of the SM meson four momentum to the axion four momentum. This approximation also ignores the interference effects from different axion-meson contributions to the production. Hence, future improvement in the axion flux prediction would be highly desirable.
With these caveats in mind, we now estimate the flux of axions produced only via meson mixing.

The mixing production is primarily driven by axion mixing with the Standard Model mesons $\pi$, $\eta$, and $\eta'$. In a convenient choice of basis, the corresponding mixing angles are given by~\cite{Bauer:2017ris,Ertas:2020xcc,Kelly:2020dda},
\begin{equation}\label{eq:mixing}
\begin{aligned}
\theta_{a \pi} & = \dfrac{1}{6} \dfrac{f_\pi}{f_a} \dfrac{m_a^2}{m_a^2-m_\pi^2}, \hspace{1 cm}
\theta_{a \eta} &= \dfrac{1}{\sqrt{6}} \dfrac{f_\pi}{f_a} \dfrac{m_a^2- 4 m_\pi^2/9}{m_a^2-m_\eta^2}, \hspace{1 cm}
\theta_{a \eta'} &= \dfrac{1}{2 \sqrt{3}} \dfrac{f_\pi}{f_a} \dfrac{m_a^2 - 16 m_\pi^2/9}{m_a^2-m_{\eta^\prime}^2}.
\end{aligned}
\end{equation}
Using \texttt{Pythia8}~\cite{Bierlich:2022pfr,Sjostrand:2014zea}, we determine how many mesons are produced per proton collision by turning on the ``{\tt SoftQCD:all = on}'' option. This is dependent on the energy of the incoming beam. 
We show the results below for the 8 GeV Booster Neutrino Beam (BNB)~\cite{Stancu:2001cpa}, 120 GeV NuMI proton beam, 400 GeV SPS beam and finally 14 TeV LHC,
\begin{equation}
	\begin{aligned}
		N_{\rm axions} = N_{\rm POT} \times  
		\begin{cases}
			0.82 |\theta_{a \pi}|^2  + 0.072 |\theta_{a \eta}|^2 + 0.0038 |\theta_{a \eta^\prime}|^2~~{\rm for~}8~ {\rm GeV~BNB} \\
			2.9 |\theta_{a \pi}|^2  + 0.33 |\theta_{a \eta}|^2  + 0.034 |\theta_{a \eta^\prime}|^2~~{\rm for~}120~ {\rm GeV~NuMI~beam} \\
			4.0 |\theta_{a \pi}|^2  + 0.46 |\theta_{a \eta}|^2  + 0.049 |\theta_{a \eta^\prime}|^2~~{\rm for~}400~ {\rm GeV~SPS~beam} \\
			 33 |\theta_{a \pi}|^2  + 3.8 |\theta_{a \eta}|^2  + 0.48 |\theta_{a \eta^\prime}|^2~~{\rm for~}14~ {\rm TeV~LHC}
		\end{cases} \, ,
	\end{aligned}
\end{equation}
where $N_{\rm POT}$ is the number of protons on target (POT). Following their production, axions travel a macroscopic distance to reach the detector and can subsequently give a displaced decay signal.

\subsection{Detection}

The produced axions will travel to the detector and decay to various final states. We will show prospects for SBND, ICARUS, DUNE, SHiP, and FASER~2 on detecting these decays. We specifically search for the signatures associated with the dimuon final states. LArTPCs identify muons by the profile of the energy loss per unit track length $dE/dx$, as a function of the residual range, defined as the distance to the particle's stopping point. 
The pair of muons will leave either one or two distinct minimally ionizing tracks depending on the opening angles and the detector's angular resolution. The SHiP experiment has a dedicated muon detector composed of three layers of scintillators with iron absorbers in between and is located at the most downstream location~\cite{Tosi:2019xok}, before which the hadron calorimeter would stop most of the hadrons. The FASER~2 experiment utilizes an emulsion detector~\cite{Nakamura:2005xov} with silver bromide crystals dispersed in gelatin media, and muons are identified by the track lengths~\cite{FASER:2019dxq}.
Among these detectors, the magnetic field in DUNE, SHiP, FASER~2, and potentially also ICAURS~\cite{Antonello:2013ypa}, would play an important role in increasing the separation between the two muon tracks. Correspondingly, the invariant mass of the dimuon system can serve as a useful handle in background mitigation.

In Table~\ref{tab:detctors}, we summarize the detector specifications. Here $E_p$ refers to the energy of the proton beam on target, $N_{\rm POT}$ is the total number of protons on target, $d$ is the distance from the target to the detector, $L$ is the length of the detector along the beam direction, while $w \times h$ is the width and height of each detector. 

In deriving the experimental prospects, we will approximate any rectangular cross section by a disk with radius shown as $r_{\rm eff}$, while FASER~2 has a circular cross section $r = 1$ m. DUNE near detector (ND) complex consists of a gaseous argon and a liquid argon time projection chamber along the beam direction; both chambers have length 5 m, while the gaseous (liquid) argon chamber has a circular (rectangular) cross section of (effective) radius ($r_{\rm eff} = 2.6$ m) $r = 2.5$ m. ICARUS has two modules of the same size next to each other; therefore, the effective cross section is multiplied by 2 in the table. 
For ICARUS, we will show the results for the protons from both the off-axis, 120 GeV NuMI beam and the default on-axis, 8 GeV BNB since the former has a better prospect for our search.
For SBND we will show only the results for 8 GeV BNB, which may achieve a better reach compared to the NuMI beam.\footnote{We thank Ornella Palamara for providing the relevant beam and luminosity information for the off-axis usage of these experiments.}
While we will not show the results for MicroBooNE explicitly, we have checked that it performs similarly to the other SBN experiments in our analysis.

\begin{table}[]
\begin{tabular}{|c|c|c|c|c|c|}
\hline
 Experiments          & $E_p$ (GeV) & $N_{\rm POT}$                                                                                            & $d$ (m) & $L$ (m)    & $w \times h$ (m$\times$m) \\ \hline
DUNE~\cite{DUNE:2020ypp,DUNE:2020fgq}       & 120         & $1.47 \times 10^{22}$                                                                                    & 574     & 5 + 5         & $7 \times 3$ ~({$r_{\rm eff} = $ 2.6 m)}            \\ \hline
SBND~\cite{Bonesini:2022pwy} $\begin{dcases} {\rm BNB} \\ {\rm NuMI} \end{dcases}$      & \makecell[c]{8 \\ 120}           & \makecell[c]{$6.6 \times 10^{20}$ \\ $3 \times 10^{21}$}     & \makecell[c]{110 \\ 410}     & 5          & $4 \times 4$ ~({$r_{\rm eff} = $ 2.3 m)}            \\ \hline
MicroBooNE~\cite{MicroBooNE:2017tkp, Bonesini:2022pwy} $\begin{dcases} {\rm BNB} \\ {\rm NuMI} \end{dcases}$ & \makecell[c]{8 \\ 120}           & \makecell[c]{$1.32 \times 10^{21}$ \\ $3 \times 10^{21}$}                                                                                  & \makecell[c]{470 \\ 685}     & 10.4      & $2.6 \times 2.3$ ~({$r_{\rm eff} = $ 1.4 m)}        \\ \hline
ICARUS~\cite{Bonesini:2022pwy}   $\begin{dcases} {\rm BNB} \\ {\rm NuMI} \end{dcases}$   & \makecell[c]{8 \\ 120}          & \makecell[c]{$6.6 \times 10^{20}$ \\ $2.5 \times 10^{21}$}                                                                                    & \makecell[c]{600 \\ 790}     & 19.9       & $(3.9 \times 3.6) \times 2$ ~({$r_{\rm eff} = $ 2.1 m)}        \\ \hline
SHiP~\cite{SHiP:2021nfo}       & 400         & $2 \times 10^{20}$                                                                                   & 70 & 50         & $5 \times 10$  ~({$r_{\rm eff} = $ 4.0 m)}          \\ \hline
FASER~2~\cite{FASER:2019aik,Feng:2022inv}    & 14000       & \makecell[c]{$1.1 \times 10^{16}$ (LHC Run 3)\\ $2.2 \times 10^{17}$ (HL-LHC)} & 480     & 5          & $r = 1$ m    \\ \hline
\end{tabular}
\caption{Detector specifications for various experiments. 
For DUNE, we have combined the lengths of the gaseous multi-purpose detector (5~m) and LArTPC (5~m). SBND, MicroBooNE, and ICARUS can serve as useful facilities as off-axis detectors with respect to the NuMI beam, along with the default on-axis BNB. Therefore, we record the locations and geometries for both beams.
} 
\label{tab:detctors}
\end{table}

\begin{figure}[]
    \centering
    \includegraphics[width=0.495\textwidth]{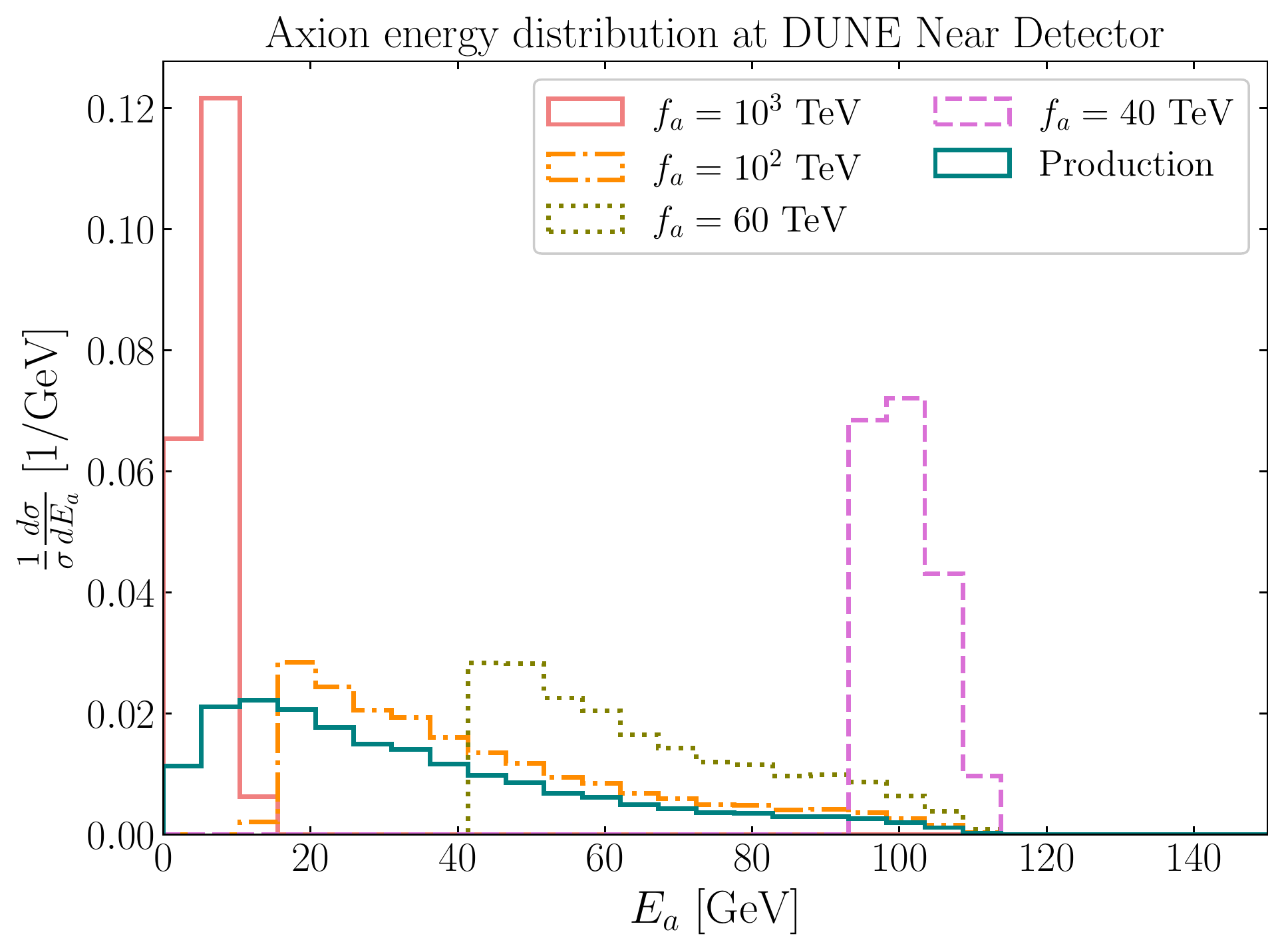}
    \includegraphics[width=0.495\textwidth]{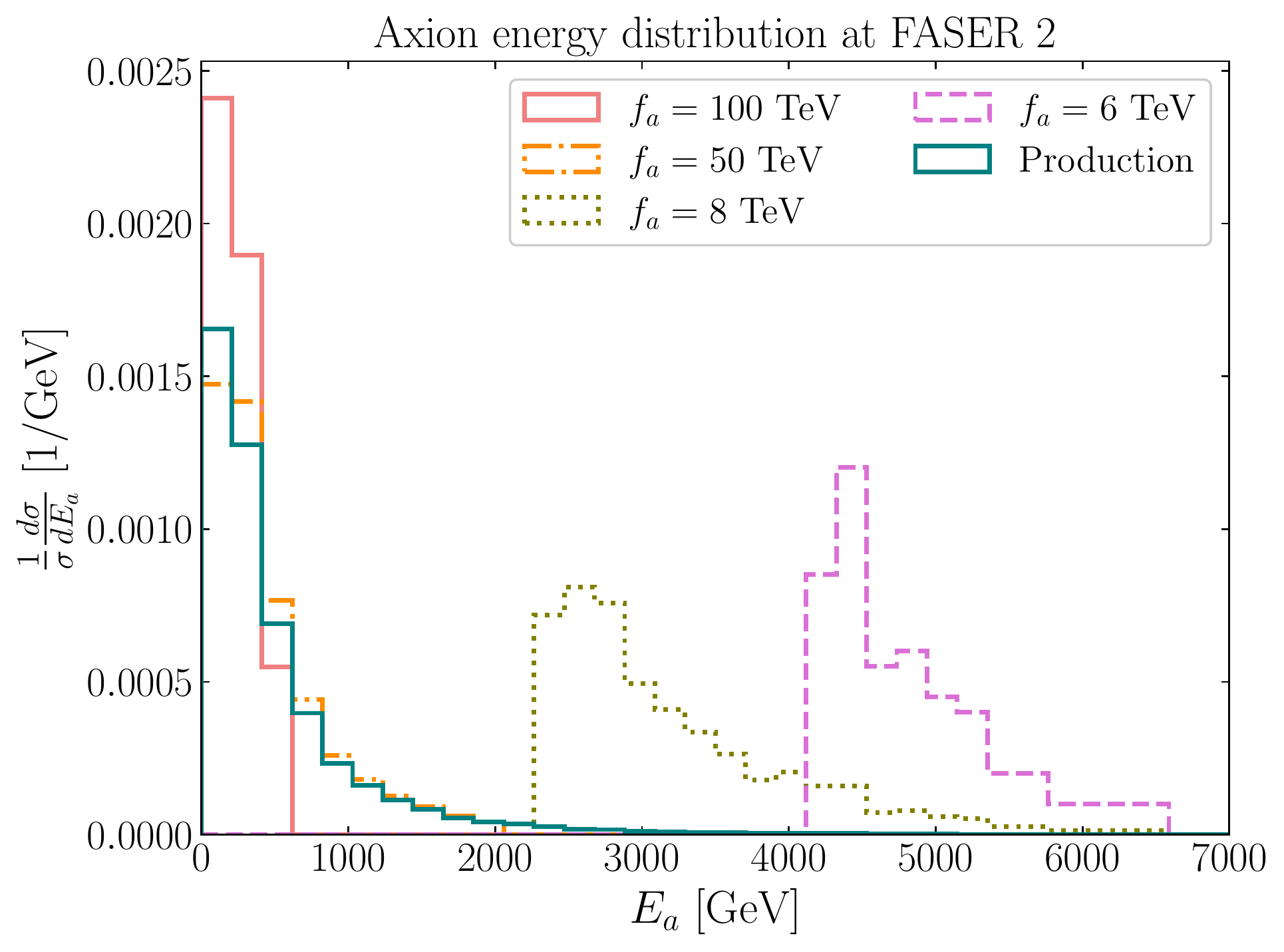}
    \caption{The differential energy distribution of the axions at production (in teal) and at the detectors (for various values of $f_a$ in other colors and line styles). We fix $c_\ell = 1/36$ and $m_a=0.8$~GeV.}
    \label{fig:ax_energy_dune}
\end{figure}

Using the above details for axion production and detector dimensions, we can compute the energy distributions of axions both at the production point and at the detector.
In Fig.~\ref{fig:ax_energy_dune} we show these results for the DUNE ND (left) and FASER~2 (right).
For both the panels, we see that the spectrum at production peaked at low energies as expected, with the axions at FASER~2 being more energetic than at DUNE ND due to larger incoming beam energies.
The axions that decay inside the detector need sufficient energy so that their lab-frame lifetime is long enough to reach the detector.
We see this explicitly in Fig.~\ref{fig:ax_energy_dune}---for smaller $f_a$, the rest-frame lifetime is shorter, and correspondingly, the axions need to be more energetic with a large Lorentz factor $\gamma$.
This is why the distributions move towards higher energies as $f_a$ decreases.

\subsection{Enhanced Effective Detector Length}\label{sec:enhanced_decay}

The axion to dimuon decay channel not only has the feature of being clean and highly identifiable but also can allow for an enhancement in the decay volume. These high-energy muons can penetrate air and rocks without significant loss in energy and deflection. In this section, we take the geometry of DUNE ND as an example and show the interplay between enhanced decay volume and signal efficiency, with their dependence on the axion mass and energy.

\begin{figure}
    \centering
    \includegraphics[width=0.7\textwidth]{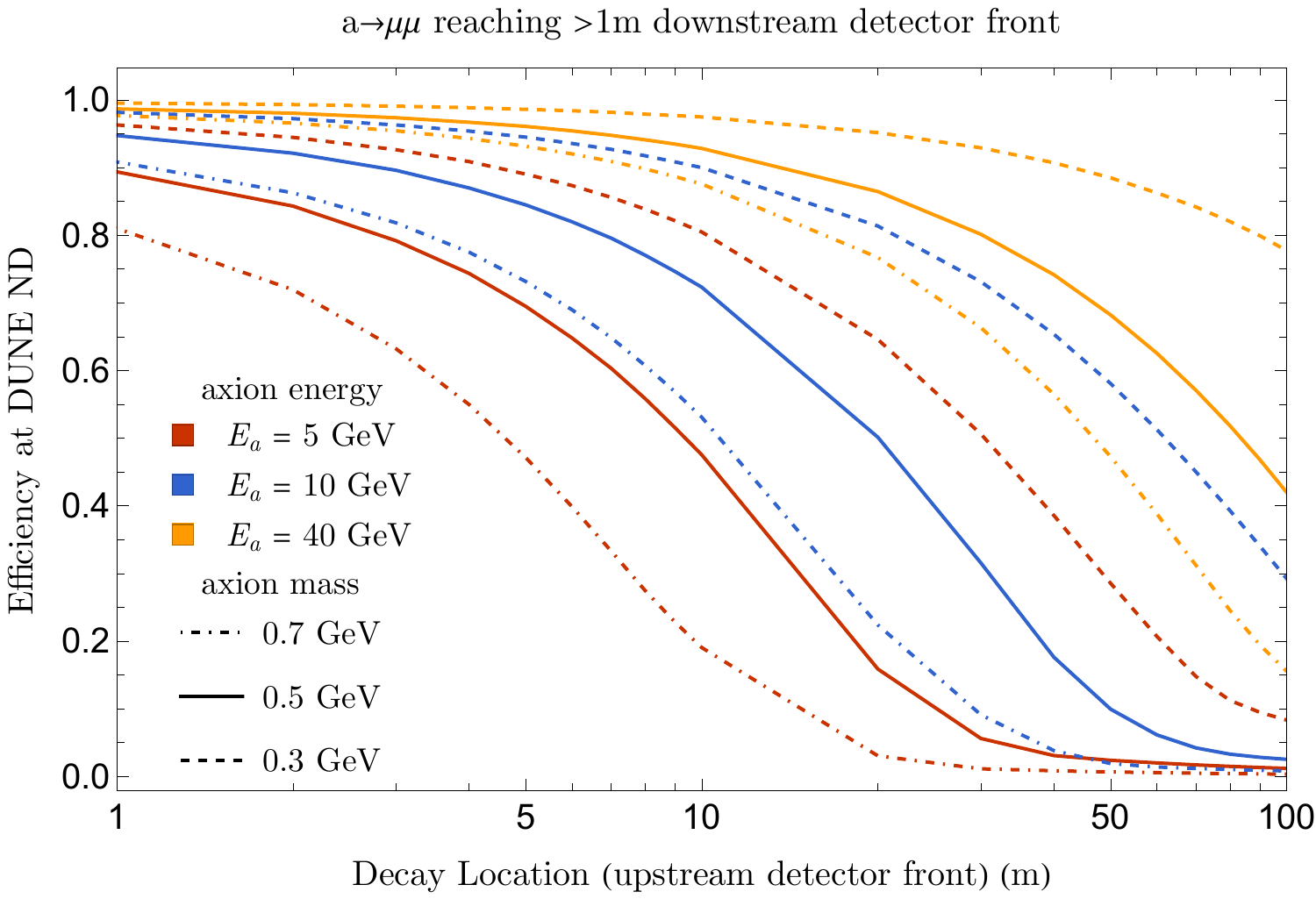}
    \includegraphics[width=0.495\textwidth]{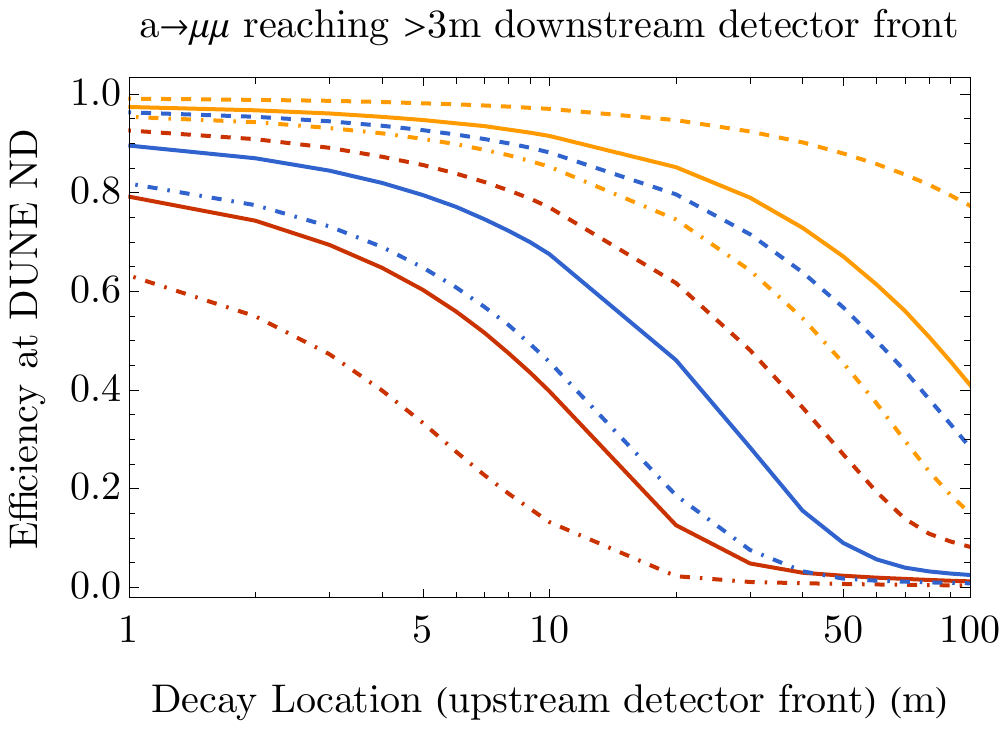}
    \includegraphics[width=0.495\textwidth]{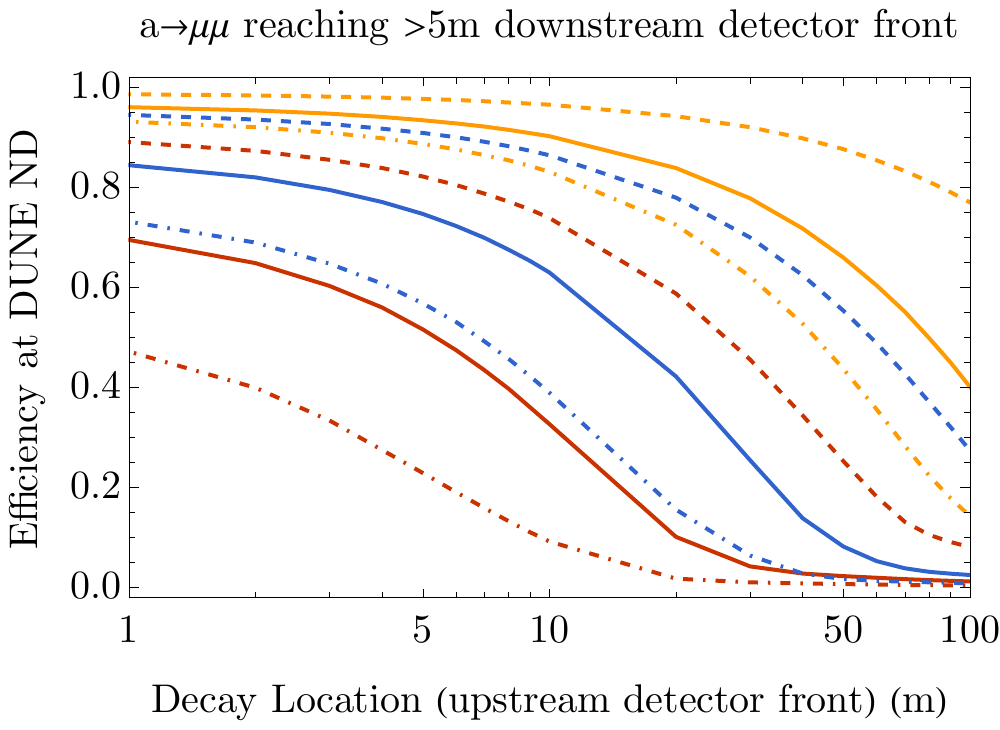}
    \caption{The signal axion decay geometrical acceptance, as a function of decay location before the front panel of the detector (upstream) for various axion masses and energies. Depending on the background rejection requirement and the actual performance of the detector, one might require a certain track length for the dimuon signals. As most of the axions are boosted, the decay is forward, so most of the signal will penetrate the whole detector. We show the acceptance fraction by requiring {\it both} the muons to arrive at least 1 meter (upper panel), which is equivalent to a track length of more than 1 meter, 3 meters (lower left panel), and 5 meters (lower right panel). The red, blue, and yellow curves correspond to axion energy of 5, 10, and 40 GeV, respectively. The dashed, solid, and dot-dashed curves correspond to axion masses of 0.3, 0.5, and 0.7 GeV. 
    }
    \label{fig:decaybefore}
\end{figure}

For axion decay before the detector, several physics effects must be considered. One is geometric acceptance: the further the decay location away from the detector, the lower the geometric acceptance. This is because the geometric acceptance would be proportional to the area divided by the distance. Thanks to the boosted kinematics of the beam dump experiments and the forward physics at colliders, the penalty is reduced by the fact that axions are already focused towards the target after getting boosted from an isotropic distribution in their rest frame.  
The second factor one needs to consider is the change in the signal-to-background ratio. Because the decay location is outside the detector, the vertex cannot be directly reconstructed within the detector volume and hence typically would have larger uncertainties in its location. In this sense, other muons, such as those generated through various nuclear interactions of the beam remnants can contribute to the background. To avoid complexities, we conservatively require both of the muons from the axion decay to arrive at the detector. The third factor is that the muons lose energy and slightly change their momentum direction while passing through matter. One can estimate the typical energy loss and deflection for the minimally ionizing muons and see that for typical high-energy muons from the axion decays with tens of GeV of energies passing through 100 meters of dirt, these effects are small~\cite{ParticleDataGroup:2020ssz}.\footnote{See Ref.~\cite{Harnik:2019zee} for approximate formulae for the energy loss and angle deflection with the charge parameter $\epsilon=1$ for muons.} Again being conservative, we do not consider the decay location more than 100 meters before the detector front panel.

Each experiment is conducted in a different setup; the amount of shielding varies, and the material before the detector varies (they can be air, dirt, or even another detector). However, the effective gain of the decay volume in the muon mode is a common feature. To concretely demonstrate this effective volume increase, we use the DUNE ND setup and simulate axion decay acceptance as a function of axion mass, energy, and muon track reconstruction requirement (track length) and show them in Fig.~\ref{fig:decaybefore}. As expected, the acceptance decreases as one increases the separation between the axion decay location and detector location. Still, for most signals, one can achieve more than 50\% acceptance even for decays occurring 30 meters before the detector. We can see that decays from heavier axions have lower acceptance for fixed axion energy. These observations lead to our discussion and presentation of projected sensitivities in the following subsection.

\subsection{Results}
In this subsection, we show the results of our projection study and our re-interpreted limits from existing experiments on heavy axions in the mass range under consideration.

\subsubsection{Future Projections}

For future projections, we analyze several neutrino experiments: SBND, ICARUS, and DUNE, as well as collider experiments searching for long-lived particles: FASER~2 and SHiP. The main idea is that the axion production would happen at the beam dump or at the primary interaction point. Subsequently, the axion would travel to the decay volume of the detector and dominantly decay into a pair of muons. For our final projections, we will assume a 90\% reconstruction efficiency of the muons along with a negligible background.

\begin{figure}
    \centering
    \includegraphics[width=0.495\textwidth]{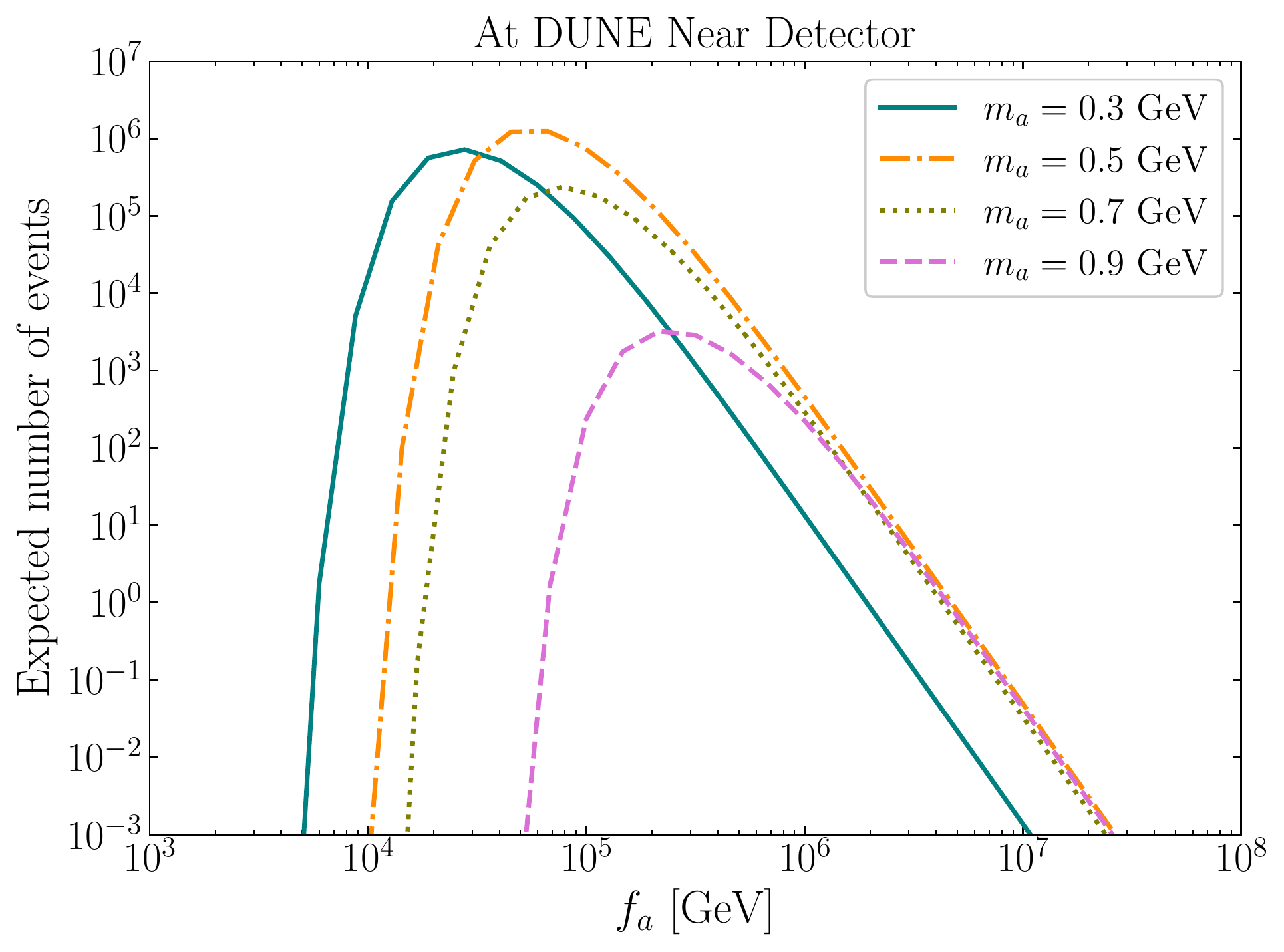}
    \includegraphics[width=0.495\textwidth]{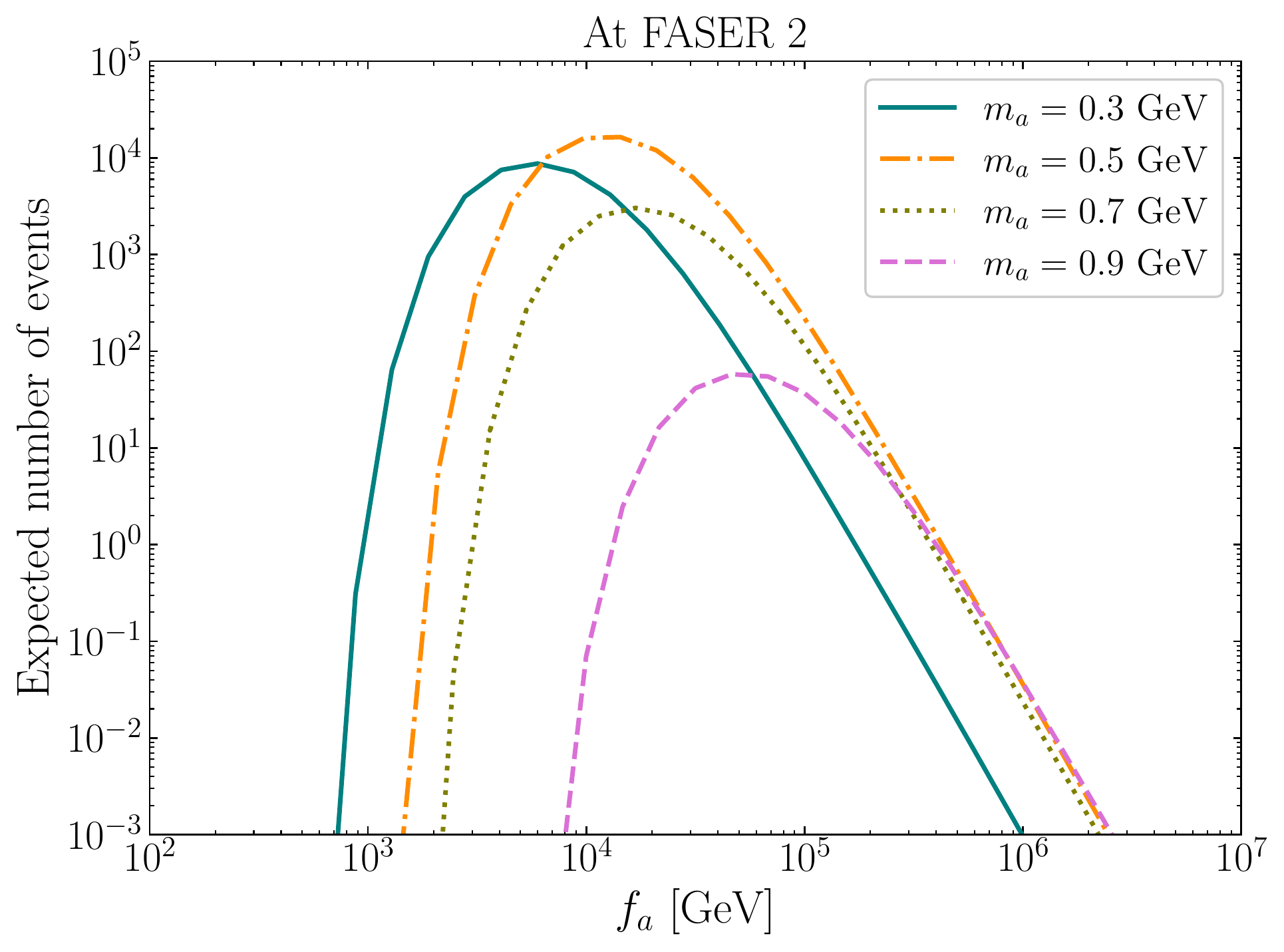}
    \includegraphics[width=0.495\textwidth]{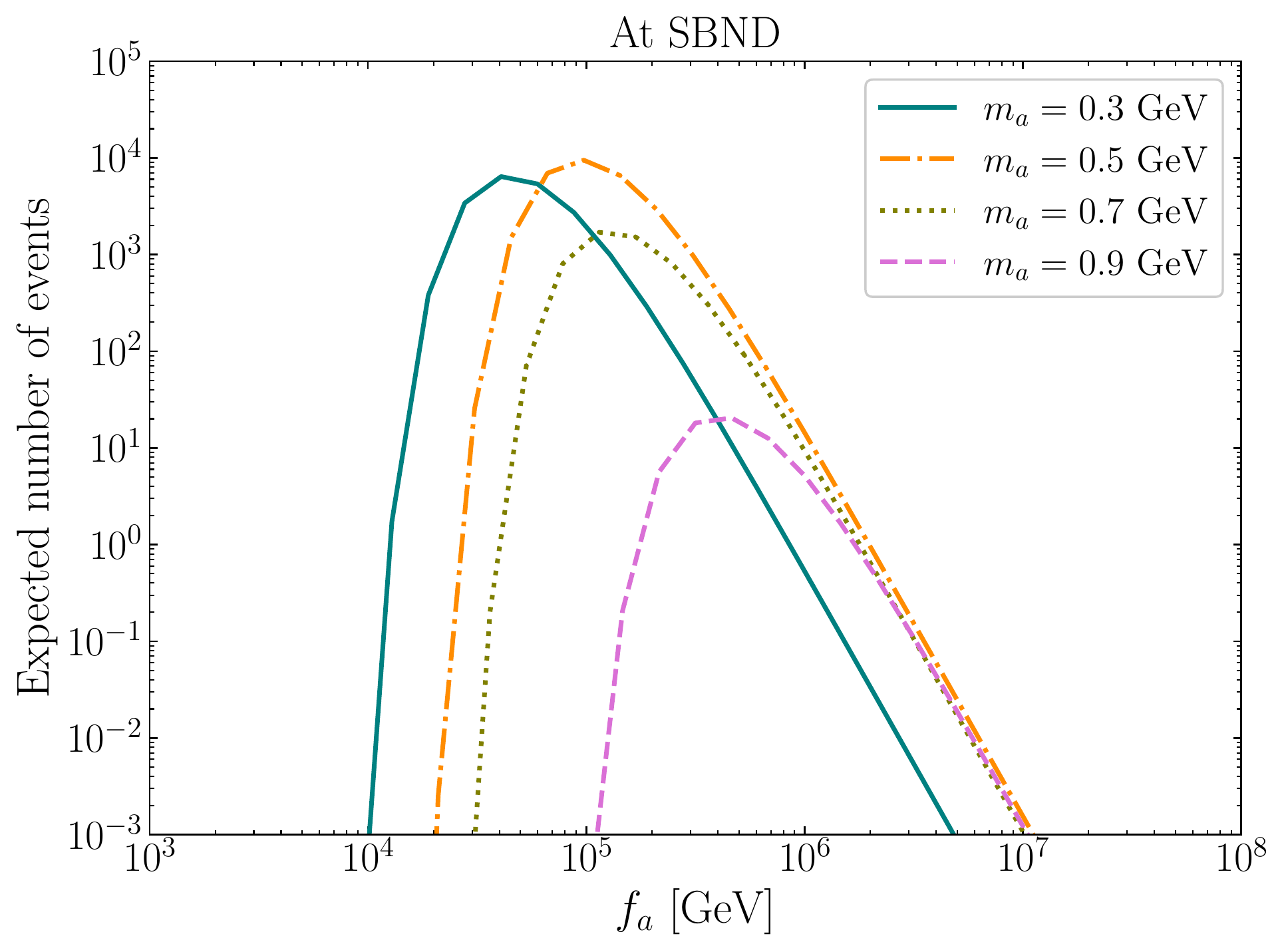}
    \includegraphics[width=0.495\textwidth]{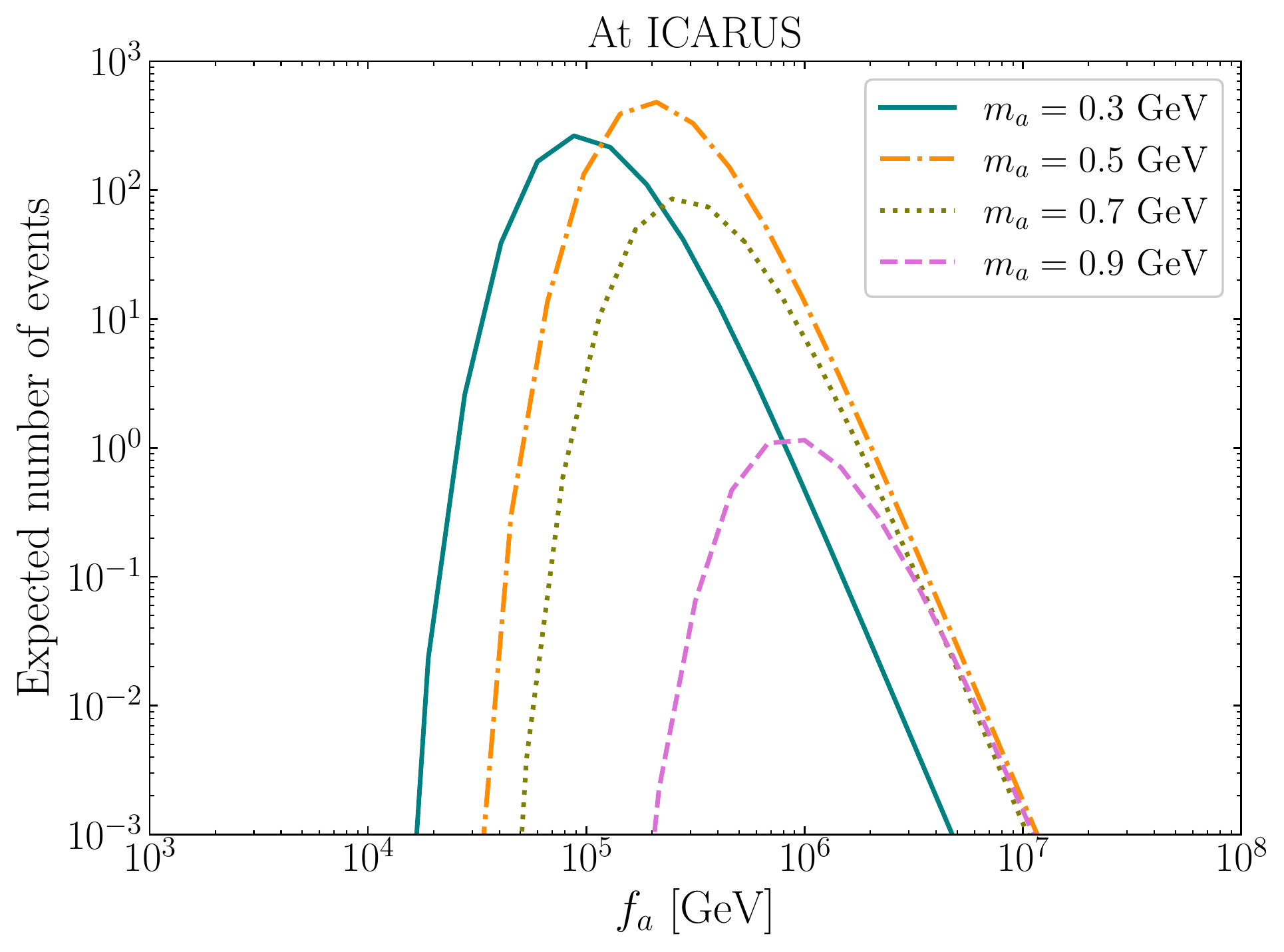}
    \caption{The expected number of events at DUNE ND, FASER~2, SBND, and ICARUS in which the axion decays into dimuons within the detector, as a function of the decay constant $f_a$. Different curves are for different axion masses $m_a$ as labeled. We fix $c_\ell = 1/36$.}
    \label{fig:flux}
\end{figure}

While an actual $\mu^+\mu^-$ background would be reducible, in LArTPCs charged pions can mimic muons, giving rise to background events. In Ref.~\cite{ArgoNeuT:2022mrm}, such backgrounds were mitigated via matching track information in ArgoNeuT with MINOS near detector, along with restrictions on track length. This latter restriction helped since pions tend to produce shorter tracks compared to muons at energies relevant to that search. 
A similar approach could be taken up at DUNE with the help of its multipurpose near detector to mitigate such backgrounds. 
Furthermore, in the case of magnetized detectors, the invariant mass would serve as a powerful discriminator as well. Dedicated simulations by experimental collaborations would be necessary to settle these aspects conclusively. 
Keeping this in mind, we show results for signal flux at various detectors in Fig.~\ref{fig:flux}, for several choices of $m_a$.
This can then be used to obtain more conservative projections for a search with non-negligible number of background events.

To this end, we first compute the number of signal events produced at the detectors of interest. For some benchmark choices of $m_a$ we show the results for SBND, ICARUS, DUNE ND, and FASER~2 in Fig.~\ref{fig:flux} as a function of $f_a$. The number of axions reaching the detector can be roughly estimated as\footnote{In our numerical study we included detailed resonance effects as discussed in Sec.~\ref{sec:production}.}
\begin{align}\label{eq:Nax}
    N_{\rm axions} \sim N_{\rm POT} \times N_{\rm mesons} \times \left(\frac{f_\pi}{f_a}\right)^2 \times {\rm BR} \times {\rm acceptance} \times \left(e^{-\frac{d}{\gamma\beta c\tau}} - e^{-\frac{d+L}{\gamma\beta c\tau}}\right).
\end{align}
Here $N_{\rm mesons}$ is the number of mesons produced per collision with each meson mixing into axions with a probability $(f_\pi/f_a)^2$, away from any resonance.
A fraction, denoted by `acceptance', of the produced axions are in the direction of detector which decay inside it with a probability $\left(e^{-\frac{d}{\gamma\beta c\tau}} - e^{-\frac{d+L}{\gamma\beta c\tau}}\right)$ into dimuons with a branching ratio `BR'. Here $\tau$ is the rest-frame lifetime of the axions, $\gamma$ is the Lorentz factor, and $\beta$ is the dimensionless velocity. 
Based on Eq.~\eqref{eq:Nax}, we see for very small values of $f_a$, although more axions are produced via the mixing factor $(f_\pi/f_a)^2$, the axions are too short-lived to reach the detector, $\gamma\beta c\tau \ll d$. 
As a result, the number of signal events gets exponentially suppressed. 
As $\gamma\beta c\tau \sim d$, the signal turns on, and eventually for large enough $f_a$, $\gamma\beta c\tau \gg d$.
In this large lifetime regime, the signal decreases as $N_{\rm axions} \propto L/f_a^4$, as can be seen via Fig.~\ref{fig:flux} and Eq.~\eqref{eq:Nax}.

\begin{figure}
    \centering
    \includegraphics[width=0.7\textwidth] {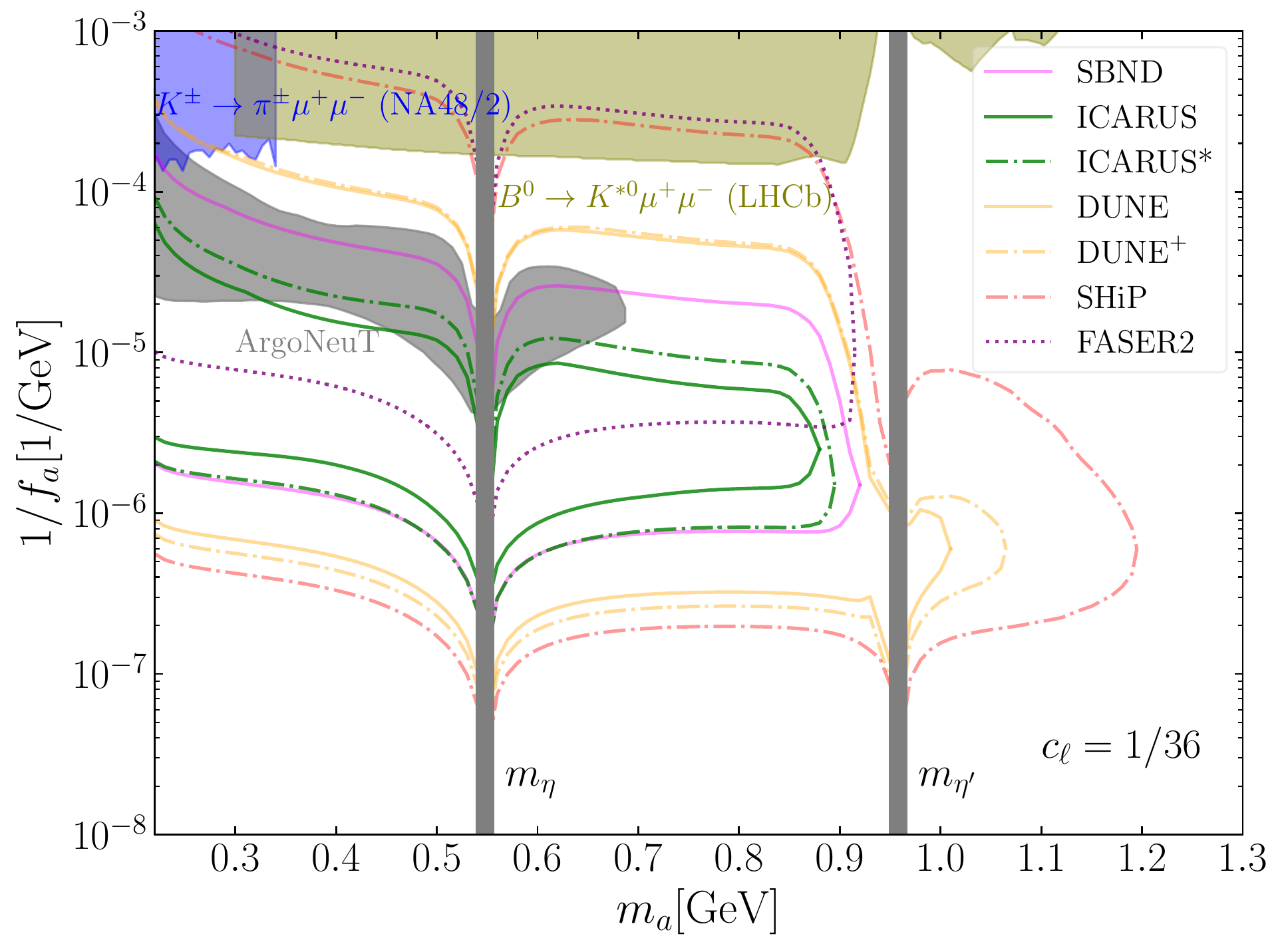}
    \includegraphics[width=0.7\textwidth] {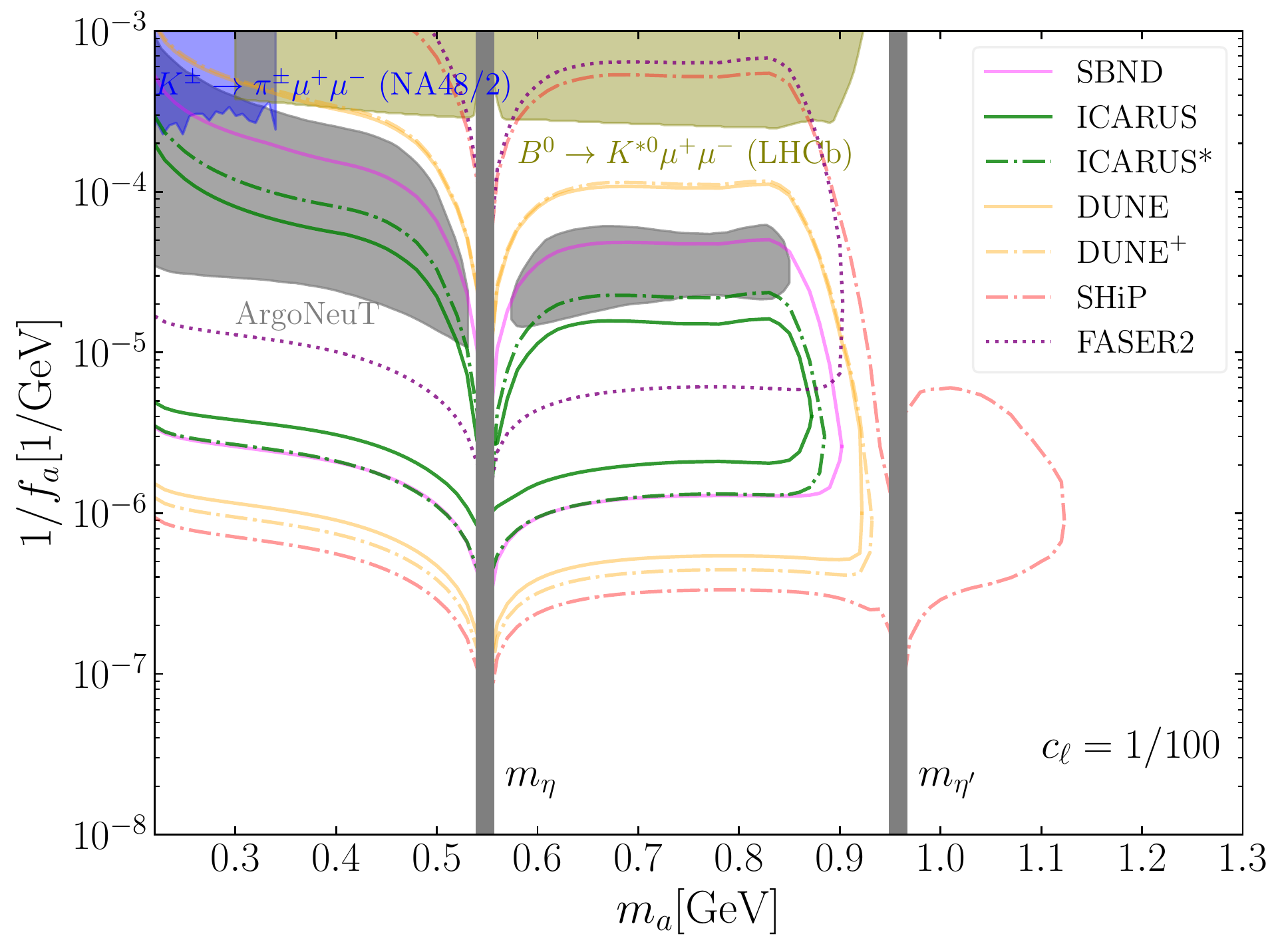}
    \caption{Constraints (shaded-regions) and projections (curves) for $c_\ell=1/36$ (upper panel) and $c_\ell=1/100$ (lower panel). The curve labeled ${\rm DUNE}^{+}$ denotes a scenario in which the axion can decay up to 30~m before the DUNE ND. The high-energy muons produced via these decays can still reach the detector. The curve labeled ${\rm ICARUS}^*$ is for the reach ICARUS would have through the off-axis NuMI beam. See text for more details.
    }
    \label{fig:reach}
\end{figure}

To derive the maximal sensitivity, we show the reach curves on the axion parameter space assuming 3 signal events for various experiments in Fig.~\ref{fig:reach}. 
Similar reach curves requiring a larger number of signal events, in the event of non-negligible backgrounds, can be obtained from Fig.~\ref{fig:flux}.
We see that among the neutrino experiments, DUNE ND would be able to provide the strongest probe, thanks to its large $N_{\rm POT}$ and decay volume. However, overall, SHiP would provide the best sensitivity. This is due to its larger beam energy, proximity to the target, as well as larger decay volume. 
While ICARUS with its on-axis BNB would constrain novel parameter space, it can operate even more powerfully via the off-axis NuMI beam
due to a larger beam energy which is not completely compensated by a larger distance. 
This is labeled as ICARUS* in Fig.~\ref{fig:reach}.
In the context of DUNE ND, we also show the result of including a distance of 30~m before the front detector panel as part of the decay volume, as discussed in Sec~\ref{sec:enhanced_decay}.
This is labeled as `DUNE$^+$' in Fig.~\ref{fig:reach}. 
While in the small lifetime, {\it i.e.}, small $f_a$ regime, the gain is not significant, in the large lifetime regime the sensitivity to $f_a$ increases as $\propto L^{1/4}$, as expected from Eq.~\eqref{eq:Nax} above.

\subsubsection{Existing Constraints}
The strongest constraints on the parameter space come from some of the rare meson decay searches which we now describe.
\paragraph{NA48/2 $K^{\pm}\rightarrow \pi \mu\mu$ search~\cite{NA482:2016sfh}:} In the presence of the axion-gluon coupling, flavor changing couplings get induced at loop-level. For example, an axion-gluon coupling induces flavor-diagonal couplings to quarks, which then can give rise to flavor-changing processes through the CKM couplings. As a result, processes such as $K\rightarrow \pi a$ with $a$ decaying into $\mu^+\mu^-$ can take place. We use the results of Refs.~\cite{Bauer:2021mvw, Bauer:2021wjo} to first compute $K\rightarrow \pi a$ amplitude. Subsequently, we use the results of Ref.~\cite{NA482:2016sfh}, where upper limits on $\text{BR}(K^{\pm}\rightarrow \pi^{\pm}X(\mu^+ \mu^-))$ were derived for various lifetimes of an unstable resonance $X$, to arrive at the constraints shown in blue in Fig.~\ref{fig:reach}.  
 
\paragraph{LHCb $B^0\rightarrow K^{*0}\mu\mu$ search~\cite{LHCb:2015nkv}:} Above the kinematic threshold of $(m_K-m_\pi)$, rare $B$ decays become the dominant probe of our scenario. Similar as above, we use the results of Ref.~\cite{Bauer:2021wjo} to first compute the rate for $B^0\rightarrow K^{*0} a$ amplitude, and then use the upper limits on $\text{BR}(B^{0}\rightarrow K^{*0}X(\mu^+ \mu^-))$ from Ref.~\cite{LHCb:2015nkv} for different lifetimes of $X$, to arrive at the constraints shown in olive in Fig.~\ref{fig:reach}. A similar bound was also derived in Ref.~\cite{Freytsis:2009ct} using rare $B$-decays for DFSZ-style axion couplings.

\paragraph{Other searches:} Searches for $K\rightarrow \pi + {\rm inv.}$~\cite{NA62:2021zjw} are also applicable when axion decays invisibly due to its large lifetime. For our scenario, however, this constraint is applicable only for $m_a < 2m_\mu$, since otherwise, axion would promptly decay into dimuons. Searches for exotic $B$ decays such as $B\rightarrow K \eta_X (\eta \pi \pi)$~\cite{BaBar:2008rth} where $\eta_X$ is one of the excited states of $\eta,\eta'$ mesons, can also be used to constrain processes such as $B\rightarrow K a (\eta \pi \pi)$~\cite{Chakraborty:2021wda}. These are mostly subdominant in our parameter space and are not shown in Fig.~\ref{fig:reach}. A recent CMS search~\cite{CMS:2021sch} for long-lived particles decaying into muon pairs can also be relevant in a similar parameter space covered by the LHCb search for $B^0\rightarrow K^{*0}\mu^+\mu^-$. This would require a dedicated recasting as appropriate for gluon-coupled axions, such as production simulation using gluon fusion and applying their kinematic selection criteria, and we do not carry that out here.

Along with these constraints from rare meson decay searches, we show in the gray-shaded region the recent ArgoNeuT constraint~\cite{ArgoNeuT:2022mrm} that the ArgoNeuT collaboration and the present authors obtained.
In this case, following their production in the beam dump, the axions would travel to the ArgoNeuT detector and dominantly decay into a pair of muons.
We consider only the axions produced by the protons that reach the hadron absorber located 318 m upstream of ArgoNeuT, which is about $13\%$ of the total $N_{\rm POT}$ collected during its run. 
The axions produced by the rest of the protons that hit the target located 1033 m upstream of the ArgoNeuT detector give a subdominant sensitivity compared to the absorber-produced axions we consider.
By searching for such muon pairs, with information from both the ArgoNeuT and MINOS near detector, we could place new constraints as shown in Fig.~\ref{fig:reach}.

\section{Examples of UV Completion}\label{sec:UV}
In this section, we describe some examples of UV models to show how the axion couplings in Eqs.~\eqref{eq:lag_gauge} and \eqref{eq:lag_lepton} can arise. We pay particular attention to models that generate the small axion-lepton couplings, compared to gauge boson couplings, considered in the previous section.

A particularly simple possibility is where $c_\ell$ is zero at tree-level but radiatively induced by the coupling of the axion with the gauge bosons. The contributions to the axion-lepton coupling from one-loop radiative corrections with $c_1, c_2$, and $c_3$ are given in Ref.~\cite{Bauer:2020jbp} for $f_a = 1~\TeV$ as
\begin{align}
    c_{\ell} 
    \simeq \left( 0.05 c_1  + 0.22 c_2 + 0.37 c_3 \right) \times 10^{-3} .
\end{align}
The coefficients depend on the renormalization scale but may only increase by a factor of $\mathcal{O}(3)$ for $f_a \simeq \mathcal{O} (10^7)~\GeV$.  Therefore, to obtain $c_\ell \simeq 1/100$ that is of interest in the previous section, one needs $c_{1,2} = \mathcal{O}(10)$, and therefore a small hierarchy from $c_3$ is necessary since $c_3$ is conventionally normalized to unity for QCD axions. This is the case if there are more heavy fermions charged under $U(1)_Y$ and $SU(2)_L$ that generate $c_{1,2}$ than those under $SU(3)_c$ for $c_3$. Some examples of enhancing certain axion couplings this way include the Kim-Nilles-Peloso mechanism~\cite{Kim:2004rp} and the clockwork axion~\cite{Kaplan:2015fuy}.

To generate an axion-lepton coupling larger than the one-loop contributions, we now describe a model based on Ref.~\cite{Buen-Abad:2021fwq}, where the coupling arises from the mixing of SM leptons with new heavy fermions. The model has a SM singlet PQ scalar $\Phi$, heavy vector-like leptons $L_i,L^c_i, E_i,E^c_i$, and heavy vector-like quarks $Q,Q^c$. The index $i=1,2,3$ runs over the SM flavor indices. The charge assignments for the left-handed fields are as given in Table~\ref{tab:charge_assignment}.
\renewcommand{\arraystretch}{1.2}
\begin{table}[ht]
\centering
\begin{tabular}{c|c|c|c|c}
\hline
\hline
Field & $SU(3)_c$ & $SU(2)_L$ & $U(1)_Y$ & $U(1)_{\rm PQ}$ \\
\hline
$L_i$ & 1  & $\Box$ & +1/2 & $-$1\\
$L^c_i$ & 1 & $\Box$ & $-$1/2 & +1\\
$E_i$ & 1  & 1 & $-$1 & $-$1\\
$E^c_i$ & 1  & 1 & +1 & +1\\
$Q$ & $\Box$  & $\Box$ & $+1/\sqrt{2}$ & $-$1/2\\
$Q^c$ & $\bar{\Box}$  & $\Box$ & $-1/\sqrt{2}$ & $-$1/2\\
$\Phi$ & 1  &1 & 0 & +1 \\
\hline
$l_i$ & 1  & $\Box$ & $-$1/2 & 0 \\
$e_i$ & 1 & 1 & +1 & 0 \\
\hline
\end{tabular}
\caption{Charge assignment for the UV model}
\label{tab:charge_assignment}
\end{table}

With this choice, the Lagrangian can be written as
\begin{align}
\mathcal{L}\supset y_{L_i} L_i l_i \Phi + y_{E_i} E_i e_i\Phi + y_Q Q Q^c\Phi + 
M_L L_i L^c_i + M_E E_i E_i^c +\rm{h.c.}.
\end{align}
Here $M_L$ and $M_E$ are vector-like masses for heavy leptons. The PQ scalar $\Phi$ has a potential suitable for symmetry breaking and we parametrize the axion as $\Phi = ((f_a+\rho)/\sqrt{2})e^{ia/f_a}$. For $M_L,M_E\gg f_a$, we can integrate out the heavy leptons at tree level to obtain their equations of motion,
\begin{align}
L^c_i = -\frac{y_{L_i}}{M_L}l_i\Phi,~~ E^c_i = - \frac{y_{E_i}}{M_E}e_i\Phi.    
\end{align}
Substituting this into the kinetic term for $L^c_i$, we get
\begin{align}
i L_i^{c\dagger} \bar{\sigma}^\mu \partial_\mu L^c_i  \rightarrow i \dfrac{|y_{L_i}|^2}{M_L^2}  l_i^\dagger \Phi^\dagger \bar{\sigma}_\mu \partial_\mu(l_i\Phi) \supset i \dfrac{|y_{L_i}|^2}{M_L^2} l_i^\dagger \Phi^\dagger \bar{\sigma}_\mu l_i \partial_\mu \Phi +\cdots.
\end{align}
Here in the last relation, we have dropped a correction to the SM fermion kinetic term.
We can carry out the above steps for $e_i$ as well. Consequently, in the EFT below $M_L, M_E$, we have
\begin{align}
\mathcal{L}\supset i \dfrac{|y_{L_i}|^2}{M_L^2} l_i^\dagger \Phi^\dagger \bar{\sigma}_\mu l_i \partial_\mu \Phi +  i \dfrac{|y_{E_i}|^2}{M_E^2} e_i^\dagger \Phi^\dagger \bar{\sigma}_\mu e_i \partial_\mu \Phi + y_Q Q Q^c\Phi +\cdots.
\end{align}
Below the scale $f_a$, we can integrate out the radial mode $\rho$ to arrive at,
\begin{align}
\mathcal{L}\supset - \frac{1}{2} \dfrac{|y_{L_i}|^2 f_a}{M_L^2}  l_i^\dagger \bar{\sigma}_\mu l_i \partial_\mu a - \frac{1}{2} \dfrac{|y_{E_i}|^2 f_a}{M_E^2} e_i^\dagger \bar{\sigma}_\mu e_i \partial_\mu a+ \frac{1}{\sqrt{2}} y_Q f_a Q Q^c e^{ia/f_a} +\cdots.   
\end{align}
We can remove the axion in the phase of the heavy quark mass term by doing a rotation, $Q \rightarrow Q e^{-ia/(2f_a)}, Q^c \rightarrow Q^c e^{-ia/(2f_a)}$ which generates the axion-gauge boson couplings via the anomaly.
With the specified quantum numbers for $Q,Q^c$, these couplings correspond to $c_1=c_2=c_3=1$ in Eq.~\eqref{eq:lag_gauge}. 
Thus the axion couplings after integrating out $Q, Q^c$ are given by,
\begin{align}
\mathcal{L}\supset - \frac{1}{2} \dfrac{|y_{L_i}|^2 f_a}{M_L^2}  l_i^\dagger \bar{\sigma}_\mu l_i \partial_\mu a - \frac{1}{2} \dfrac{|y_{E_i}|^2 f_a}{M_E^2} e_i^\dagger \bar{\sigma}_\mu e_i \partial_\mu a+ \frac{\alpha_s}{8\pi f_a}a G\tilde{G} +\frac{\alpha_2}{8\pi f_a}a W\tilde{W} + \frac{\alpha_1}{8\pi f_a}a B\tilde{B}.      
\end{align}
We can write this in terms of the Dirac spinors to relate to Eq.~\eqref{eq:lag_lepton},
\begin{align}
c_\ell = \frac{1}{2}\left(\dfrac{|y_{L_i}|^2 f_a^2}{M_L^2} + \dfrac{|y_{E_i}|^2 f_a^2}{M_E^2}\right).
\end{align}
In the context of this UV completion, we see that the leptonic coupling of the axion can naturally be suppressed, consistent with our assumption of $f_a \ll M_L, M_E$ for integrating out the heavy leptons. This then serves as an example for our choice of $c_\ell = 1/36, 1/100 \ll 1$ as the benchmark leptonic couplings.

\section{Conclusion and Outlook}\label{sec:concl}

The QCD axion is a well-motivated extension of the Standard Model that can address the strong {\it CP} problem and other open questions in astrophysics and cosmology. A heavy QCD axion is especially appealing from the viewpoint of the axion quality problem. In this work, we analyze the discovery potential of the heavy QCD axions in neutrino experiments and long-lived particle searches. We exploit the unique signals and low backgrounds associated with the dimuon final states. We focus on models where the axion is coupled to leptons in addition to its universal coupling to gauge bosons, which can arise in a large class of axion models. The axion mass range of interest is above the dimuon threshold, and below the GeV scale, so the decay to dimuons is the dominant channel. 

Assuming negligible background, which we argue might be achieved by additional measurements of particle track lengths and invariant masses, we show projected sensitivity curves for FASER~2, SHiP, DUNE, SBND, and ICARUS in Fig.~\ref{fig:reach}. Both DUNE and SHiP provide the best sensitivity for probing large decay constants $f_a \simeq 10^{6\mathchar`-7}~\GeV$ depending on the leptonic coupling constant $c_\ell$. Larger decay constant leads to overly suppressed production rates. On the other hand, FASER~2 and SHiP offer the best opportunities in probing smaller decay constants $f_a \simeq 10^{3\mathchar`-4}~\GeV$. Smaller decay constants result in axion decays too rapid for the axions to reach the detectors, but such a small $f_a$ is anyway excluded by existing constraints from rare meson decay searches. Lastly, the coverage for SBND and ICARUS centers around $f_a \simeq 10^{4.5\mathchar`-6}~\GeV$. These promising projections call for dedicated simulations by the experimental collaborations to estimate the background and place important constraints on the motivated parameter space of heavy QCD axions.

\section*{Acknowledgements}
We thank Roni Harnik, Kevin Kelly, Simon Knapen, Ornella Palamara for useful discussions. S.K. was supported in part by the National Science Foundation (NSF) grant PHY-1915314 and the U.S. Department of Energy Contract DE-AC02-05CH11231. S.K. thanks the Mainz Institute of Theoretical Physics of the Cluster of Excellence PRISMA+ (Project ID 39083149) for its hospitality while this work was in progress. The work was supported in part by the U.S. Department of Energy (DOE) under grant No. DE-SC0022345 (Z.L.) and DE-SC0011842 (R.C.). Z.L. thanks the Aspen Center for Physics, where part of this work was completed, which is supported by National Science Foundation (NSF) grant PHY-1607611. Z.L. acknowledges the support of the Munich Institute for Astro-, Particle and BioPhysics (MIAPbP), where part of this work was conducted, which was funded by the Deutsche Forschungsgemeinschaft (DFG, German Research Foundation) under Germany's Excellence Strategy-EXC2094-390783311.

\bibliographystyle{utphys}
\bibliography{refs}

\end{document}